\documentclass[12pt]{article}

\RequirePackage{epsf}
\RequirePackage{longtable}

\begin{document}

\title{The Cosmological Constant}

\author{Sean M. Carroll \\
        Enrico Fermi Institute and Department of Physics \\
        University of Chicago \\ 
        5640 S.~Ellis Ave. \\
        Chicago, IL 60637 USA \\
        e-mail: carroll@theory.uchicago.edu \\
        http://pancake.uchicago.edu/{\~{}}carroll/}

\date{}
\maketitle

\begin{abstract}
This is a review of the physics and cosmology of the cosmological
constant.  Focusing on recent developments, I present a 
pedagogical overview of cosmology in the presence of a
cosmological constant, observational constraints on its
magnitude, and the physics of a small (and potentially nonzero) 
vacuum energy.

\end{abstract}


\begin{center}
Submitted to {\sl Living Reviews in Relativity}, December 1999
\end{center}

\vfill

\noindent astro-ph/00004075 \\ EFI-2000-13

\newpage

\section{Introduction}
\label{section:introduction}

\subsection{Truth and beauty}
\label{section:truth}

Science is rarely tidy.  We ultimately seek a unified explanatory
framework characterized by elegance and simplicity; along the way,
however, our aesthetic impulses must occasionally
be sacrificed to the desire to encompass the largest possible range 
of phenomena ({\it i.e.}, to fit the data).  
It is often the case that an otherwise compelling theory, in order
to be brought into agreement with observation, 
requires some apparently unnatural modification. 
Some such modifications may eventually be
discarded as unnecessary once the phenomena are better understood;
at other times, advances in our theoretical understanding will
reveal that a certain theoretical compromise is only superficially 
distasteful, when in fact it arises as the consequence of a
beautiful underlying structure.

General relativity is a paradigmatic example of a scientific theory of
impressive power and simplicity.  The cosmological constant,
meanwhile, is a paradigmatic example of a modification, originally
introduced \cite{einstein}
to help fit the data, which appears at least on the surface
to be superfluous and unattractive.  Its original role, to allow
static homogeneous solutions to Einstein's equations in the presence
of matter, turned out to be unnecessary when the expansion of the
universe was discovered \cite{hubble}, 
and there have been a number of subsequent
episodes in which a nonzero cosmological constant was put forward as
an explanation for a set of observations and later withdrawn when the
observational case evaporated.  Meanwhile, particle theorists
have realized that the cosmological constant can be interpreted as a
measure of the energy density of the vacuum.  This energy
density is the sum of a number of apparently unrelated contributions,
each of magnitude much larger than the upper limits on the
cosmological constant today; the question of why the observed vacuum
energy is so small in comparison to the scales of particle physics has
become a celebrated puzzle, although it is usually thought to be
easier to imagine an unknown mechanism which would set it precisely to
zero than one which would suppress it by just the right amount to
yield an observationally accessible cosmological constant.

This checkered history has led to a certain reluctance to consider 
further invocations of a nonzero cosmological constant; however,
recent years have provided the best evidence yet that this elusive
quantity does play an important dynamical role in the universe.  
This possibility, although still far from a certainty, makes it
worthwhile to review the physics and astrophysics of the cosmological
constant (and its modern equivalent, the energy of the vacuum).

There are a number of other reviews of various aspects of the
cosmological constant; in the present article I will outline
the most relevant issues, but not
try to be completely comprehensive, focusing instead on 
providing a pedagogical introduction and explaining recent
advances.  For astrophysical aspects, I did not try to 
duplicate much of the material in Carroll, Press
and Turner \cite{cpt}, which should be consulted
for numerous useful formulae and a discussion of several kinds of
observational tests not covered here.  Some earlier discussions
include \cite{felten,charlton,sandage88}, and subsequent reviews include
\cite{cohn,sahni,Turner:1998}.  The classic discussion of the 
physics of
the cosmological constant is by Weinberg \cite{weinberg}, with
more recent work discussed by \cite{cohn,sahni}.  For introductions
to cosmology, see \cite{kt,lindebook,peebles}.

\subsection{Introducing the cosmological constant}
\label{section:introducing}

Einstein's original field equations are
\begin{equation}
  R_{\mu\nu} - {1\over 2}Rg_{\mu\nu} 
  = 8\pi GT_{\mu\nu}\ .
  \label{einstein}
\end{equation}
(I use conventions in which $c=1$, and will also set $\hbar=1$
in most of the formulae to follow, but Newton's constant
will be kept explicit.)  On very large scales the universe is 
spatially homogeneous
and isotropic to an excellent approximation, which implies that
its metric takes the Robertson-Walker form
\begin{equation}
  {\rm d}s^2 = -{\rm d}t^2 + a^2(t)R_0^2\left[
  {{{\rm d}r^2}\over{1-kr^2}} + r^2 {\rm d}\Omega^2\right]\ ,
  \label{rwmetric}
\end{equation}
where ${\rm d}\Omega^2 = {\rm d}\theta^2 + \sin^2\theta {\rm d}
\phi^2$ is the metric on a two-sphere.  The curvature parameter
$k$ takes on values $+1$, $0$, or $-1$ for positively curved,
flat, and negatively curved spatial sections, respectively.
The scale factor characterizes the relative size of the
spatial sections as a function of time; we have written it in
a normalized form $a(t) = R(t)/R_0$, where 
the subscript $0$ will always refer to a quantity evaluated
at the present time.
The redshift $z$ undergone by radiation from a comoving object
as it travels to us today is related to the scale factor
at which it was emitted by
\begin{equation}
  a = {1\over {(1+z)}}\ .
\end{equation}

The energy-momentum sources may be modeled as a perfect fluid,
specified by an energy density $\rho$ and isotropic pressure $p$
in its rest frame.  The energy-momentum tensor of such a fluid
is
\begin{equation}
  T_{\mu\nu} = (\rho +p)U_\mu U_\nu + p g_{\mu\nu}\ ,
  \label{tmunufluid}
\end{equation}
where $U^\mu$ is the fluid four-velocity.
To obtain a Robertson-Walker solution to Einstein's equations,
the rest frame of the fluid must be that of a comoving observer
in the metric (\ref{rwmetric}); in that case, Einstein's
equations reduce to the two Friedmann equations
\begin{equation}
  H^2 \equiv \left({{\dot a}\over a}\right)^2 =
  {{8\pi G}\over 3}\rho - {k\over{a^2R_0^2}}\ ,
  \label{feq1}
\end{equation}
where we have introduced the Hubble parameter
$H\equiv \dot a /a$, and
\begin{equation}
  {{\ddot a}\over a} = -{4\pi G \over 3}(\rho + 3p)\ .
  \label{feq2}
\end{equation}

Einstein was interested in finding static ($\dot a = 0$)
solutions, both due to his hope that general relativity would
embody Mach's principle that matter determines inertia,
and simply to account for the astronomical data as they
were understood at the time.\footnote{This account gives
short shrift to the details of what actually happened;
for historical background see \cite{weinberg}.}  A static universe
with a positive energy density is compatible with
(\ref{feq1}) if the spatial curvature is positive ($k=+1$)
and the density is appropriately tuned; however, 
(\ref{feq2}) implies that $\ddot a$ will never vanish in
such a spacetime if the pressure $p$ is also nonnegative
(which is true for most forms of matter, and certainly
for ordinary sources such as stars and gas).  Einstein
therefore proposed a modification of his equations, to
\begin{equation}
  R_{\mu\nu} - {1\over 2}Rg_{\mu\nu} 
  + \Lambda g_{\mu\nu}
  = 8\pi GT_{\mu\nu}\ , 
  \label{einsteinl}
\end{equation}
where $\Lambda$ is a new free parameter, the cosmological 
constant.  Indeed, the left-hand side of (\ref{einsteinl}) 
is the most general local, coordinate-invariant,
divergenceless, symmetric, two-index tensor we can
construct solely from the metric and its first and second
derivatives.  With this modification, the Friedmann
equations become
\begin{equation}
  H^2 = {{8\pi G}\over 3}\rho + {\Lambda\over 3} 
  - {k\over{a^2R_0^2}}\ .
  \label{feq1l}
\end{equation}
and
\begin{equation}
  {{\ddot a}\over a} = -{4\pi G \over 3}(\rho + 3p) + 
  {\Lambda \over 3} \ .
  \label{feq2l}
\end{equation}
These equations admit a static solution with positive spatial
curvature and all the parameters $\rho$, $p$, and $\Lambda$
nonnegative.  This solution is called the ``Einstein static
universe."

The discovery by Hubble that the universe is expanding eliminated
the empirical need for a static world model (although the
Einstein static universe continues to thrive in the toolboxes
of theorists, as a crucial step in the construction of conformal
diagrams).  It has also been criticized on the grounds that
any small deviation from a perfect balance between the 
terms in (\ref{feq2l}) will rapidly grow into a runaway
departure from the static solution.

Pandora's box, however, is not so easily closed.  The
disappearance of the original motivation for introducing the
cosmological constant did not change its status as a
legitimate addition to the gravitational field equations,
or as a parameter to be constrained
by observation.  The only way to purge $\Lambda$ from
cosmological discourse would be to measure all of the
other terms in (\ref{feq1l}) to sufficient precision 
to be able to conclude that the $\Lambda/3$
term is negligibly small in comparison, a feat which has
to date been out of reach.  As discussed below, there
is better reason than ever before to believe that
$\Lambda$ is actually nonzero, and Einstein may not
have blundered after all.

\subsection{Vacuum energy}
\label{section:vacuumenergy}

The cosmological constant $\Lambda$ is a dimensionful
parameter with units of (length)$^{-2}$.  From the point
of view of classical general relativity, there is no
preferred choice for what the length scale defined by
$\Lambda$ might be.  Particle physics, however, brings
a different perspective to the question.  The cosmological
constant turns out to be a measure of the energy density
of the vacuum --- the state of lowest energy --- and although
we cannot calculate the vacuum energy with any
confidence, this identification allows us to consider the
scales of various contributions to the cosmological
constant \cite{zeld,bludman}.

Consider a single scalar field $\phi$, with potential
energy $V(\phi)$.  The action can be written
\begin{equation}
  S = \int d^4x\, \sqrt{-g}\left[ {1\over 2} g^{\mu\nu}
  \partial_\mu\phi \partial_\nu\phi - V(\phi)\right]
\end{equation}
(where $g$ is the determinant of the metric tensor 
$g_{\mu\nu}$), and the corresponding energy-momentum tensor is
\begin{equation}
  T_{\mu\nu} = {1\over 2}\partial_\mu\phi \partial_\nu\phi
  + {1\over 2} (g^{\rho\sigma}\partial_\rho\phi \partial_\sigma\phi)
  g_{\mu\nu} - V(\phi)g_{\mu\nu}\ .
\end{equation}
In this theory, the configuration with the lowest energy 
density (if it exists) will be one in which there is no
contribution from kinetic or gradient energy, implying
$\partial_\mu\phi = 0$, for which 
$T_{\mu\nu}= - V(\phi_0)g_{\mu\nu}$, where $\phi_0$
is the value of $\phi$ which minimizes $V(\phi)$.  There
is no reason in principle why $V(\phi_0)$ should vanish.
The vacuum energy-momentum tensor can thus be written
\begin{equation}
  T^{\rm vac}_{\mu\nu} = -\rho_{\rm vac} g_{\mu\nu}\ ,
  \label{tmunuvac}
\end{equation}
with $\rho_{\rm vac}$ in this example given by $V(\phi_0)$.
(This form for the vacuum energy-momentum tensor can also be
argued for on the more general grounds that it is the only
Lorentz-invariant form for $T^{\rm vac}_{\mu\nu}$.)
The vacuum can therefore be thought of as a perfect fluid
as in (\ref{tmunufluid}), with
\begin{equation}
  p_{\rm vac} = -\rho_{\rm vac}\ .
\end{equation}
The effect of an
energy-momentum tensor of the form (\ref{tmunuvac}) is
equivalent to that of a cosmological constant, as can be
seen by moving the $\Lambda g_{\mu\nu}$ term in
(\ref{einsteinl}) to the right-hand side and setting
\begin{equation}
  \rho_{\rm vac} = \rho_\Lambda \equiv {{\Lambda}\over{8\pi G}}\ .
\end{equation}
This equivalence is the origin of the identification of
the cosmological constant with the energy of the vacuum.
In what follows, I will use the terms ``vacuum energy" and
``cosmological constant" essentially interchangeably.

It is not necessary to introduce scalar fields
to obtain a nonzero vacuum energy.  The action for general
relativity in the presence of a ``bare'' cosmological
constant $\Lambda_0$ is
\begin{equation}
  S = {1\over 16\pi G}\int d^4x\, \sqrt{-g} (R -
  2\Lambda_0)\ ,
  \label{action}
\end{equation}
where $R$ is the Ricci scalar.  Extremizing this action
(augmented by suitable matter terms)
leads to the equations (\ref{einsteinl}).  Thus, the cosmological
constant can be thought of as simply a constant term in
the Lagrange density of the theory.  Indeed, (\ref{action})
is the most general covariant action we can construct out of 
the metric and its first and second derivatives, and is
therefore a natural starting point for a theory of gravity.

Classically, then, the effective cosmological constant is
the sum of a bare term $\Lambda_0$ and the potential energy
$V(\phi)$, where the latter may change with time as the
universe passes through different phases.  Quantum
mechanics adds another contribution, from the
zero-point energies associated with vacuum fluctuations.
Consider a simple harmonic oscillator,
{\it i.e.}\ a particle moving in a one-dimensional potential
of the form $V(x)={1\over 2}\omega^2 x^2$.  
Classically, the ``vacuum'' for this system is the state in
which the particle is motionless and at the minimum of the
potential ($x=0$), for which the energy in this case vanishes.
Quantum-mechanically, however, the uncertainty principle
forbids us from isolating the particle both in position and
momentum, and we find that the lowest
energy state has an energy $E_0 = {1\over 2} \hbar\omega$
(where I have temporarily re-introduced explicit factors
of $\hbar$ for clarity).
Of course, in the absence of gravity either system actually
has a vacuum energy which is completely arbitrary; we could
add any constant to the potential (including, for example,
$-{1\over 2} \hbar\omega$) without changing the theory.  
It is important, however, that the zero-point energy
depends on the system, in this case on the frequency
$\omega$.

A precisely analogous situation holds in field theory.
A (free) quantum field can be thought of as a collection of an
infinite number of harmonic oscillators in momentum space.
Formally, the zero-point energy of such an infinite collection
will be infinite.  (See \cite{weinberg,cpt} for further
details.)  If, however, we discard the very high-momentum
modes on the grounds that we trust our theory only up to
a certain ultraviolet momentum cutoff $k_{\rm max}$, we
find that the resulting energy density is of the form
\begin{equation}
  \rho_{\Lambda} \sim \hbar k_{\rm max}^4\ .
\end{equation}
This answer could have been guessed by dimensional analysis;
the numerical constants which have been neglected will depend
on the precise theory under consideration.  
Again, in the absence of gravity this energy has no effect,
and is traditionally discarded (by a process known as 
``normal-ordering'').  However, gravity does exist, and
the actual value of the vacuum energy has important
consequences.  (And the vacuum fluctuations themselves are
very real, as evidenced by the Casimir effect \cite{casimir}.)

The net cosmological constant, from this point of view, is the
sum of a number of apparently disparate contributions,
including potential energies from scalar fields and zero-point
fluctuations of each field theory degree of freedom, as well
as a bare cosmological constant $\Lambda_0$.  Unlike the last
of these, in the first two cases we can at least make
educated guesses at the magnitudes.  In the Weinberg-Salam
electroweak model, the phases of broken and unbroken
symmetry are distinguished by a potential energy 
difference of approximately $M_{\rm EW} \sim 200$~GeV
(where 1~GeV $= 1.6\times 10^{-3}$~erg); the
universe is in the broken-symmetry phase during our
current low-temperature epoch, and is believed to have
been in the symmetric phase at sufficiently high temperatures
in the early universe.  The effective cosmological constant
is therefore different in the two epochs; absent some
form of prearrangement, we would naturally expect a 
contribution to the vacuum energy today of order
\begin{equation}
  \rho_\Lambda^{\rm EW} \sim (200~{\rm GeV})^4
  \sim 3\times 10^{47} {\rm ~erg/cm}^3\ .
\end{equation}
Similar contributions can arise even without invoking
``fundamental" scalar fields.  In the strong interactions,
chiral symmetry is believed to be broken by a nonzero
expectation value of the quark bilinear $\bar q q$ (which 
is itself a scalar, although constructed from fermions).
In this case the energy difference between the symmetric
and broken phases is of order the QCD scale $M_{\rm QCD}
\sim 0.3$~GeV, and we would expect a corresponding
contribution to the vacuum energy of order
\begin{equation}
  \rho_\Lambda^{\rm QCD} \sim (0.3~{\rm GeV})^4
  \sim 1.6\times 10^{36} {\rm ~erg/cm}^3\ .
\end{equation}
These contributions are joined by those from any number
of unknown phase transitions in the early universe,
such as a possible contribution from grand unification
of order $M_{\rm GUT}\sim 10^{16}$~GeV.
In the case of vacuum fluctuations, we should choose our
cutoff at the energy past which we no longer trust our
field theory.  If we are confident that we can use ordinary
quantum field theory all the way up to the Planck scale
$M_{\rm Pl} = (8\pi G)^{-1/2} \sim
10^{18}$~GeV, we expect a contribution of order
\begin{equation}
 \rho_\Lambda^{\rm Pl} \sim (10^{18}~{\rm GeV})^4
  \sim 2\times 10^{110} {\rm ~erg/cm}^3\ .
  \label{planckrho}
\end{equation}
Field theory may fail earlier, although quantum gravity
is the only reason we have to believe it will fail at
any specific scale.

As we will discuss later, cosmological observations imply 
\begin{equation}
  |\rho_\Lambda^{\rm (obs)}| \leq (10^{-12}~{\rm GeV})^4
  \sim 2\times 10^{-10} {\rm ~erg/cm}^3\ ,
  \label{obsrho}
\end{equation}
much smaller than any of the individual effects listed
above.  The ratio of (\ref{planckrho}) to (\ref{obsrho})
is the origin of the famous discrepancy of 120 orders
of magnitude between the theoretical and observational
values of the cosmological constant.
There is no obstacle to imagining that all of 
the large and apparently unrelated contributions listed add
together, with different signs, to produce a net
cosmological constant consistent with the limit (\ref{obsrho}),
other than the fact that it seems ridiculous.   We know
of no special symmetry which could enforce a vanishing
vacuum energy while remaining consistent with the
known laws of physics; this conundrum is the ``cosmological
constant problem''. 
In section \ref{section:physics}
we will discuss a number of issues related to this
puzzle, which at this point remains one of the most
significant unsolved problems in fundamental physics.

\section{Cosmology with a cosmological constant}
\label{section:cosmology}

\subsection{Cosmological parameters}
\label{section:parameters}

From the Friedmann equation
(\ref{feq1}) (where henceforth we take the effects of a 
cosmological constant into account by including the vacuum energy 
density $\rho_\Lambda$ into the total density $\rho$), for any
value of the Hubble parameter $H$ there is a critical value
of the energy density such that the spatial geometry is flat
($k=0$):
\begin{equation}
  \rho_{\rm crit} \equiv {{3H^2}\over{8\pi G}}\ .
\end{equation}
It is often most convenient to measure the total energy
density in terms of the critical density, by introducing the
density parameter
\begin{equation}
  \Omega \equiv {\rho \over {\rho_{\rm crit}}}
  =\left({{8\pi G}\over 3H^2}\right) \rho\ .
\end{equation}
One useful feature of this parameterization is a direct
connection between the value of $\Omega$ and the spatial
geometry:
\begin{equation}
  k = {\rm sgn}(\Omega-1)\ .
  \label{kandomega}
\end{equation}
[Keep in mind that some references still use ``$\Omega$''
to refer strictly to the density parameter in matter, even
in the presence of a cosmological constant; with this
definition (\ref{kandomega}) no longer holds.]

In general, the energy density $\rho$ will include
contributions from various distinct components.
From the point of view of cosmology, the relevant feature of
each component is how its energy density evolves as the universe
expands.  Fortunately, it is often (although not always) the
case that individual components $i$ have very simple equations
of state of the form
\begin{equation}
  p_i = w_i \rho_i\ ,
\end{equation}
with $w_i$ a constant.  Plugging this equation of state into
the energy-momentum conservation equation $\nabla_\mu T^{\mu\nu}
=0$, we find that the energy density has a power-law dependence
on the scale factor,
\begin{equation}
  \rho_i \propto a^{-n_i}\ ,
\end{equation}
where the exponent is related to the equation of state parameter by
\begin{equation}
  n_i = 3(1+w_i)\ .
  \label{nandw}
\end{equation}
The density parameter in each component is defined in the
obvious way,
\begin{equation}
  \Omega_i \equiv {{\rho_i}\over {\rho_{\rm crit}}}
  =\left({{8\pi G}\over 3H^2}\right) \rho_i\ ,
\end{equation}
which has the useful property that 
\begin{equation}
  {{\Omega_i}\over {\Omega_j}} \propto a^{-(n_i-n_j)}\ .
  \label{omegaprop}
\end{equation}

The simplest example of a component of this form is a set
of massive particles with negligible relative velocities,
known in cosmology as ``dust'' or simply ``matter''.
The energy density of such particles is given by their
number density times their rest mass; as the universe
expands, the number density is inversely proportional to the
volume while the rest masses are constant, yielding
$\rho_{\rm M}\propto a^{-3}$. For relativistic particles,
known in cosmology as ``radiation'' (although any relativistic
species counts, not only photons or even strictly massless
particles), the energy density is the number density times
the particle energy, and the latter is proportional to $a^{-1}$
(redshifting as the universe expands); the radiation energy
density therefore scales as $\rho_{\rm R}\propto a^{-4}$.
Vacuum energy does not change as the universe expands, so
$\rho_\Lambda \propto a^0$; from (\ref{nandw}) this implies a
negative pressure, or positive tension, when the vacuum energy is 
positive.  Finally, for some purposes it is
useful to pretend that the $-ka^{-2}R_0^{-2}$ term in (\ref{feq1})
represents an effective ``energy density in curvature'',
and define $\rho_k \equiv -(3k/8\pi GR_0^2)a^{-2}$.  We can define a
corresponding density parameter 
\begin{equation}
  \Omega_k = 1-\Omega\ ;
\end{equation}
this relation is simply (\ref{feq1}) divided by $H^2$.  Note that the
contribution from $\Omega_k$ is (for obvious reasons) not 
included in the definition of $\Omega$.  The usefulness of
$\Omega_k$ is that it contributes to the expansion rate
analogously to the honest density parameters $\Omega_i$; we
can write
\begin{equation}
  H(a) =
  H_0 \left(\sum_{i (k)} \Omega_{i0} a^{-n_i}\right)^{1/2}\ ,
  \label{hofa}
\end{equation}
where the notation $\sum_{i (k)}$ reflects the fact that the
sum includes $\Omega_k$ in addition to the various components
of $\Omega = \sum_i \Omega_i$.  The most popular equations of
state for cosmological energy sources can thus be summarized
as follows:
\begin{equation}
  \begin{tabular}{l|c|c}
   & $w_i$ & $n_i$ \\
  \hline
  matter & 0 & 3 \\
  radiation & $1/3$ & 4 \\
  ``curvature'' & $-1/3$ & 2 \\
  vacuum & $-1$ & 0 \\
  \end{tabular}
\end{equation}

The ranges of values of the $\Omega_i$'s which are allowed in
principle (as opposed to constrained by observation) will depend
on a complete theory of the matter fields, but lacking that we
may still invoke energy conditions to get a handle on what
constitutes sensible values.  The most appropriate condition
is the dominant energy condition (DEC), which
states that $T_{\mu\nu}l^\mu l^\nu \geq 0$, and
$T^\mu{}_\nu l^\mu$ is non-spacelike, for any
null vector $l^\mu$; this implies that energy does not flow faster
than the speed of light \cite{he}.  For a perfect-fluid energy-momentum
tensor of the form (\ref{tmunufluid}), these two requirements
imply that $\rho + p \geq 0$ and $|\rho| \geq |p|$, respectively.
Thus, either the density is positive and greater in magnitude
than the pressure, or the density is negative and equal in
magnitude to a compensating positive pressure; in terms of the
equation-of-state parameter $w$, we have either positive
$\rho$ and $|w|\leq 1$ or negative $\rho$ and $w=-1$.  That is,
a negative energy density is allowed only if it is in the form
of vacuum energy.  (We have actually modified the conventional
DEC somewhat, by using only null vectors $l^\mu$ rather than
null or timelike vectors; the traditional condition would 
rule out a negative cosmological constant, which there is no
physical reason to do.)

There are good reasons to believe that the energy density in
radiation today is much less than that in matter.  Photons,
which are readily detectable, contribute $\Omega_\gamma
\sim 5\times 10^{-5}$, mostly in the 
$2.73~^\circ$K  cosmic microwave background
\cite{ressellturner,fixsen,scott}.  
If neutrinos are sufficiently low mass as to
be relativistic today, conventional
scenarios predict that they contribute approximately the same
amount \cite{kt}.  
In the absence of sources which are even more exotic,
it is therefore useful to parameterize the universe today by
the values of $\Omega_{\rm M}$ and $\Omega_\Lambda$, with
$\Omega_k = 1 - \Omega_{\rm M} - \Omega_\Lambda$, keeping
the possibility of surprises always in mind.

One way to characterize a specific
Friedmann-Robertson-Walker model is by the values of
the Hubble parameter and the various energy densities
$\rho_i$.  (Of course, reconstructing the history of such a
universe also requires an understanding of the microphysical
processes which can exchange energy between the different
states.)  It may be difficult, however, to directly measure the
different contributions to $\rho$, and it is therefore useful
to consider extracting these quantities from the behavior of
the scale factor as a function of time.  A traditional measure
of the evolution of the expansion rate is the deceleration
parameter 
\begin{equation}
  \begin{array}{rcl}
  q & \equiv & \displaystyle{-{{\ddot{a}a}\over {\dot a}^2}}\cr 
  & = & {\displaystyle \sum_i {{n_i - 2}\over 2}\Omega_i} \cr
  & = & {\displaystyle {1\over 2}\Omega_{\rm M} - \Omega_\Lambda}\ ,
  \end{array}
  \label{q}
\end{equation}
where in the last line we have assumed that the universe is
dominated by matter and the cosmological constant.  
Under the assumption that $\Omega_\Lambda = 0$, measuring
$q_0$ provides a direct measurement of the current density
parameter $\Omega_{{\rm M}0}$; however, once $\Omega_\Lambda$
is admitted as a possibility there is no single parameter
which characterizes various universes, and for most purposes
it is more convenient to simply quote experimental results
directly in terms of $\Omega_{\rm M}$ and $\Omega_\Lambda$.
[Even this parameterization, of course, bears a certain theoretical
bias which may not be justified; ultimately, the only unbiased
method is to directly quote limits on $a(t)$.]

Notice that
positive-energy-density sources with $n>2$ cause the universe to
decelerate while $n<2$ leads to acceleration; the more rapidly
energy density redshifts away, the greater the tendency towards
universal deceleration.  An empty universe ($\Omega=0$, $\Omega_k = 1$)
expands linearly with time; sometimes called the ``Milne
universe'', such a spacetime is really flat Minkowski space
in an unusual time-slicing.

\subsection{Model universes and their fates}
\label{section:models}

In the remainder of this section we will explore the behavior
of universes dominated by matter and vacuum energy, $\Omega = 
\Omega_{\rm M} + \Omega_\Lambda = 1-\Omega_k$.  According to
(\ref{q}), a positive cosmological constant accelerates the
universal expansion, while a negative cosmological constant
and/or ordinary matter tend to decelerate it.  The relative
contributions of these components change with time; according
to (\ref{omegaprop}) we have
\begin{equation}
  \Omega_\Lambda \propto a^2 \Omega_k \propto
  a^3 \Omega_{\rm M}\ .
\end{equation}
For $\Omega_\Lambda < 0$, the universe will always recollapse
to a Big Crunch,
either because there is a sufficiently high matter density or
due to the eventual domination of the negative cosmological constant.
For $\Omega_\Lambda > 0$ the universe will expand forever unless
there is sufficient matter to cause recollapse before 
$\Omega_\Lambda$ becomes dynamically important.  For
$\Omega_\Lambda = 0$ we have the familiar situation in which
$\Omega_{\rm M} \leq 1$ universes expand forever and
$\Omega_{\rm M} > 1$ universes recollapse; notice, however, that
in the presence of a cosmological constant there is no necessary
relationship between spatial curvature and the fate of the universe.
(Furthermore, we cannot reliably determine that the universe will
expand forever by any set of measurements of $\Omega_\Lambda$ and
$\Omega_{\rm M}$; even if we seem to live in a parameter space
that predicts eternal expansion, there is always the possibility
of a future phase transition which could change the equation of
state of one or more of the components.)

Given $\Omega_{\rm M}$, the value of $\Omega_\Lambda$ for which
the universe will expand forever is given by
\begin{equation}
  \Omega_\Lambda \geq \left\{ \begin{array}{ll}
  0 & ~~ 0\leq\Omega_{\rm M} \leq 1 \\
  {\displaystyle 4\Omega_{\rm M}\cos^3\left[{1\over 3}\cos^{-1}\left(
  {{1-\Omega_{\rm M}}\over{\Omega_{\rm M}}}\right)
  + \frac{4\pi}{3} \right]} & ~~ \Omega_{\rm M} > 1\ .\end{array}
  \right.
\end{equation}
Conversely, if the cosmological constant is sufficiently large
compared to the matter density, the universe has always been
accelerating, and rather than a Big Bang its early history
consisted of a period of gradually slowing contraction to a 
minimum radius before beginning its current expansion.  The
criterion for there to have been no singularity in the past is
\begin{equation}
  \Omega_\Lambda \geq 
  4\Omega_{\rm M}{\rm coss}^3\left[{1\over 3}{\rm coss}^{-1}\left(
  {{1-\Omega_{\rm M}}\over{\Omega_{\rm M}}}\right)\right]\ ,
\end{equation}
where ``coss'' represents $\cosh$ when $\Omega_{\rm M} < 1/2$, and
$\cos$ when $\Omega_{\rm M} > 1/2$.

\begin{figure}[t]
  \epsfxsize = 8cm
  \centerline{\epsfbox{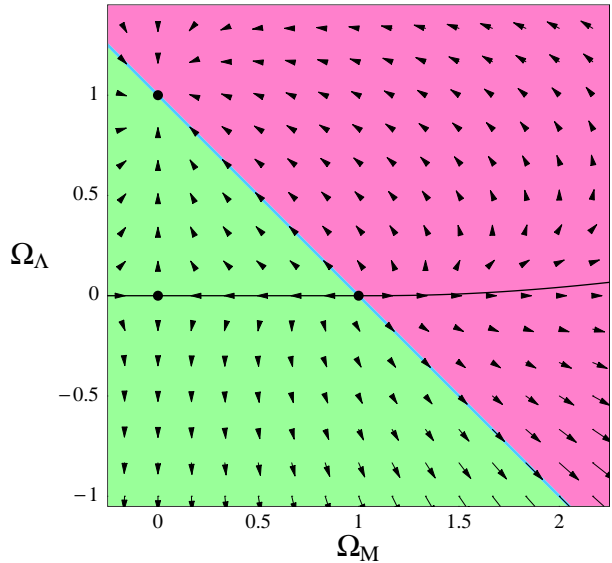}}
  \caption{Dynamics for $\Omega = \Omega_{\rm M} + \Omega_\Lambda$.
  The arrows indicate the direction of evolution of the 
  parameters in an expanding
  universe.}
  \label{fig:ovecpale}
\end{figure}

The dynamics of universes with $\Omega = \Omega_{\rm M} +
\Omega_\Lambda$ are summarized in Figure (\ref{fig:ovecpale}), in
which the arrows indicate the evolution of these parameters in
an expanding universe.  (In a contracting universe they would
be reversed.)
This is not a true phase-space plot, despite the superficial
similarities.  One important difference is that a universe passing
through one point can pass through the same point again but moving
backwards along its trajectory, by first going to infinity and
then turning around (recollapse).

Figure (\ref{fig:ovecpale}) includes three fixed points, at
$(\Omega_{\rm M}, \Omega_\Lambda)$ equal to $(0, 0)$, $(0, 1)$,
and $(1, 0)$.  The attractor among these at $(1, 0)$ is known
as de~Sitter space --- a universe with no matter density, dominated
by a cosmological constant, and with scale factor growing 
exponentially with time.  The fact that this point is an attractor
on the diagram is another way of understanding the cosmological
constant problem.  A universe with initial conditions located
at a generic point on the diagram will, after several expansion
times, flow to de~Sitter space if it began above the recollapse
line, and flow to infinity and back to recollapse if it began
below that line.  Since our universe has undergone a large 
number of $e$-folds of expansion since early times, it must 
have begun at a non-generic point in order not to have evolved
either to de~Sitter space or to a Big Crunch.  The only other
two fixed points on the diagram are the saddle point at
$(\Omega_{\rm M}, \Omega_\Lambda) = (0, 0)$, corresponding
to an empty universe, and the repulsive fixed point at
$(\Omega_{\rm M}, \Omega_\Lambda) = (1, 0)$, known as the
Einstein-de~Sitter solution.  Since our universe is not empty,
the favored solution from this combination of theoretical and
empirical arguments is the Einstein-de~Sitter universe.
The inflationary scenario 
\cite{Guth:1981,Linde:1982,Albrecht:1982} provides a mechanism
whereby the universe can be driven to the line 
$\Omega_{\rm M}+\Omega_\Lambda = 1$ (spatial flatness), so
Einstein-de~Sitter is a natural expectation if we imagine that
some unknown mechanism sets $\Lambda=0$.  As discussed below,
the observationally favored universe is located on this line
but away from the fixed points, near 
$(\Omega_{\rm M}, \Omega_\Lambda) = (0.3, 0.7)$.  It is 
fair to conclude that naturalness arguments have a somewhat
spotty track record at predicting cosmological parameters.

\subsection{Surveying the universe}
\label{section:surveying}

\begin{figure}[t]
  \epsfxsize = 8cm
  \centerline{\epsfbox{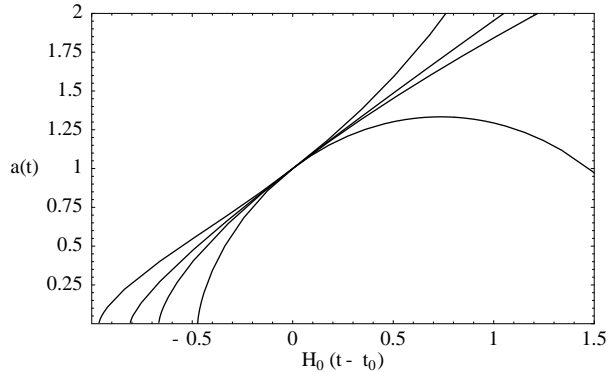}}
  \caption{Expansion histories for different values of 
  $\Omega_{\rm M}$ and $\Omega_\Lambda$.  From top to bottom,
  the curves describe $(\Omega_{\rm M}, \Omega_\Lambda) =
  (0.3, 0.7)$, $(0.3, 0.0)$, $(1.0, 0.0)$, and $(4.0, 0.0)$.}
  \label{fig:aoft}
\end{figure}

The lookback time from the present day to an object at
redshift $z_*$ is given by
\begin{equation}
  \begin{array}{rcl}
  t_0 - t_* & = & {\displaystyle \int^{t_0}_{t_*} dt }\cr
  & = & {\displaystyle \int^{1}_{1/(1+z_*)}{{da}\over{a H(a)}}}\ ,
  \end{array}
  \label{age}
\end{equation}
with $H(a)$ given by (\ref{hofa}).
The age of the universe is obtained by taking the
$z_*\rightarrow \infty$ ($t_*\rightarrow 0$) limit.  For
$\Omega = \Omega_{\rm M} = 1$, this yields the familiar answer
$t_0 = (2/3)H_0^{-1}$; the age decreases as $\Omega_{\rm M}$
is increased, and increases as $\Omega_\Lambda$ is increased.
Figure (\ref{fig:aoft}) shows the expansion history of the universe
for different values of these parameters and $H_0$ fixed; it
is clear how the acceleration caused by $\Omega_\Lambda$ leads
to an older universe.
There are analytic approximation formulas which estimate
(\ref{age}) in various regimes \cite{weinberg,kt,cpt}, but generally
the integral is straightforward to perform numerically.

In a generic curved spacetime, there is no preferred notion of
the distance between two objects.  Robertson-Walker spacetimes
have preferred foliations, so it is possible to define sensible
notions of the distance between comoving objects --- those whose
worldlines are normal to the preferred slices.  Placing ourselves
at $r=0$ in the coordinates defined by (\ref{rwmetric}), the 
coordinate distance $r$ to another comoving object is independent
of time.  It can be converted to a physical distance at any 
specified time $t_*$ by multiplying by the scale factor $R_0a(t_*)$,
yielding a number which will of course change as the universe
expands.  However, intervals along spacelike slices are not
accessible to observation, so it is typically more convenient to
use distance measures which can be extracted from observable
quantities.  These include the luminosity distance,
\begin{equation}
  d_L \equiv \sqrt{{L\over{4\pi F}}}\ ,
\end{equation}
where $L$ is the intrinsic luminosity and $F$ the measured
flux; the proper-motion distance,
\begin{equation}
  d_M \equiv {u\over{\dot \theta}}\ ,
\end{equation}
where $u$ is the transverse proper velocity and $\dot\theta$ the
observed angular velocity; and the angular-diameter distance,
\begin{equation}
  d_A \equiv {D\over \theta}\ ,
\end{equation}
where $D$ is the proper size of the object and $\theta$ its
apparent angular size.  All of these definitions reduce to the
usual notion of distance in a Euclidean space.  In a 
Robertson-Walker universe,
the proper-motion distance turns out to equal the physical
distance along a spacelike slice at $t=t_0$:
\begin{equation}
  d_M = R_0 r\ .
\end{equation}
The three measures are related by
\begin{equation}
  d_L = (1+z)d_M = (1+z)^2 d_A\ ,
\end{equation}
so any one can be converted to any other for sources of known
redshift.

The proper-motion distance between sources at redshift $z_1$ 
and $z_2$ can
be computed by using $ds^2=0$ along a light ray, where $ds^2$
is given by (\ref{rwmetric}).  We have
\begin{equation}
  \begin{array}{rcl}
  d_M(z_1, z_2) & = & R_0 (r_2 - r_1) \cr
  & = & \displaystyle{
  R_0 \, {\rm sinn}\left[\int^{t_2}_{t_1} {{dt}\over{R_0a(t)}}
  \right]} \cr
  & = & \displaystyle{
  {1\over{H_0\sqrt{|\Omega_{k0}|}}} \, {\rm sinn}\left[
  H_0\sqrt{|\Omega_{k0}|}\int^{1/(1+z_2)}_{1/(1+z_1)} 
  {{da}\over{a^2 H(a)}} \right]}\ ,
  \end{array}
  \label{dmeq}
\end{equation}
where we have used (\ref{feq1}) to solve for $R_0 = 1/(H_0
\sqrt{|\Omega_{k0}|})$, $H(a)$ is again given by (\ref{hofa}),
and ``sinn($x$)'' denotes $\sinh(x)$ when $\Omega_{k0} <0$, 
$\sin(x)$ when $\Omega_{k0} >0$, and $x$ when $\Omega_{k0}=0$.
An analytic approximation formula can be found in \cite{pen99}.
Note that, for large redshifts, the dependence of the various
distance measures on $z$ is not necessarily monotonic.

The comoving volume element in a Robertson-Walker universe
is given by
\begin{equation}
  dV = {{R_0^3 r^2}\over{\sqrt{1-kr^2}}}drd\Omega\ ,
\end{equation}
which can be integrated analytically to obtain the volume
out to a distance $d_M$:
\begin{equation}
  \begin{array}{rcl}
  V(d_M) & = & \displaystyle{
  {1\over{2H_0^{3}\Omega_{k0}}}\left[H_0d_M
  \sqrt{1+H_0^2\Omega_{k0}d_M^2}\right. }\cr
  & & \qquad\qquad \qquad\qquad \displaystyle{
  \left.  - {1\over{\sqrt{|\Omega_{k0}|}}}
  \, {\rm sinn}^{-1}(H_0\sqrt{|\Omega_{k0}|}d_M)\right]}\ ,
  \end{array}
  \label{vol}
\end{equation}
where ``sinn'' is defined as below (\ref{dmeq}).

\subsection{Structure formation}
\label{section:structure}

The introduction of a cosmological constant changes the 
relationship between the matter density and expansion rate from
what it would be in a matter-dominated universe, which in turn
influences the growth of large-scale structure.  The effect is similar
to that of a nonzero spatial curvature, and complicated by
hydrodynamic and nonlinear effects on small scales, but is
potentially detectable through sufficiently careful observations.

The analysis of the evolution of structure is greatly abetted by
the fact that perturbations start out very small (temperature
anisotropies in the microwave background imply that the density
perturbations were of order $10^{-5}$ at recombination), and
linearized theory is effective.  In this regime, the fate of 
the fluctuations is in the hands of two competing effects:
the tendency of self-gravity to make overdense regions collapse,
and the tendency of test particles in the background expansion to
move apart.  Essentially, the effect of vacuum energy is to
contribute to expansion but not to the self-gravity of
overdensities, thereby acting to suppress the growth of
perturbations \cite{kt,peebles}.  

For sub-Hubble-radius perturbations in a cold dark matter
component, a Newtonian analysis suffices.  (We may of course
be interested in super-Hubble-radius modes, or the evolution
of interacting or relativistic particles, but the simple
Newtonian case serves to illustrate the relevant physical
effect.)  If the energy density in dynamical matter
is dominated by CDM, the linearized Newtonian
evolution equation is
\begin{equation}
  \ddot\delta_{\rm M} + 2{{\dot a}\over a}\dot\delta_{\rm M}
  = 4\pi G\rho_{\rm M}\delta_{\rm M}\ .
  \label{deltaeq}
\end{equation}
The second term represents an effective frictional force due to
the expansion of the universe, characterized by a timescale
$(\dot a /a)^{-1}=H^{-1}$, while the right hand side is a forcing
term with characteristic timescale $(4\pi G\rho_{\rm M})^{-1/2}
\approx \Omega_{\rm M}^{-1/2}H^{-1}$.  Thus, when $\Omega_{\rm M}
\approx 1$, these effects are in balance and CDM perturbations
gradually grow; when $\Omega_{\rm M}$ dips appreciably below
unity (as when curvature or vacuum energy begin to dominate),
the friction term becomes more important and perturbation growth
effectively ends.  In fact (\ref{deltaeq}) can be directly
solved \cite{heath} to yield
\begin{equation}
  \delta_{\rm M}(a) = {5\over 2}H_0^2\Omega_{{\rm M}0}{{\dot a}\over a}
  \int_0^a H^{-3}(a')\, da'\ ,
\end{equation}
where $H(a)$ is given by (\ref{hofa}).  There exist analytic
approximations to this formula \cite{cpt}, as well as analytic
expressions for flat universes \cite{Eisenstein:1997}.
Note that this analysis is
consistent only in the linear regime; once perturbations on a given
scale become of order unity, they break away from the Hubble flow 
and begin to evolve as isolated systems.

\section{Observational tests}
\label{section:tests}

It has been suspected for some time now that there are good
reasons to think that a cosmology with an appreciable cosmological
constant is the best fit to what we know about the universe
\cite{Peebles:1984,Turner:1984,KS,Efstathiou:1990,Fujii:1991,
KGB,Krauss:1995,Ostriker:1995,Turner:1997}.
However, it is only very recently that the observational case
has tightened up considerably, to the extent that, as the year
2000 dawns, more experts than not believe that
there really is a positive vacuum energy exerting a measurable
effect on the evolution of the universe.  In this section I
review the major approaches which have led to this shift.

\subsection{Type Ia supernovae}
\label{section:sne}

The most direct and theory-independent way to measure the cosmological
constant would be to actually determine the value of the scale factor
as a function of time.  Unfortunately, the appearance of $\Omega_k$
in formulae such as (\ref{dmeq}) renders this difficult.  
Nevertheless, with sufficiently precise information about the 
dependence of a distance measure on redshift we can disentangle
the effects of spatial curvature, matter, and vacuum energy, and
methods along these lines have been popular ways to try to constrain
the cosmological constant.

Astronomers measure distance in terms of the ``distance modulus''
$m-M$, where $m$ is the apparent magnitude of the source and $M$
its absolute magnitude.  The distance modulus is related to the
luminosity distance via
\begin{equation}
  m - M = 5 \log_{10}[d_L({\rm Mpc})] + 25\ .
\end{equation}
Of course, it is easy to measure the apparent magnitude, but
notoriously difficult to infer the absolute magnitude of a distant
object.  Methods to estimate the relative absolute luminosities of 
various kinds of objects (such as galaxies with certain characteristics)
have been pursued,
but most have been plagued by unknown evolutionary effects or
simply large random errors \cite{sandage88}.

Recently, significant progress has been made by using Type Ia
supernovae as ``standardizable candles''.  Supernovae are rare ---
perhaps a few per century in a Milky-Way-sized galaxy --- but 
modern telescopes allow observers to probe very deeply into 
small regions of the sky, covering a very large number of galaxies
in a single observing run.  Supernovae are also bright, and Type Ia's
in particular all seem to be of nearly uniform intrinsic luminosity
(absolute magnitude $M\sim -19.5$, typically comparable to the
brightness of the entire host galaxy in which they appear)
\cite{branch}.  They
can therefore be detected at high redshifts ($z\sim 1$), 
allowing in principle a good handle on cosmological effects
\cite{tammann79,Goobar:1995}.  

\begin{figure}[t]
  \epsfxsize = 8cm
  \centerline{
  \epsfbox{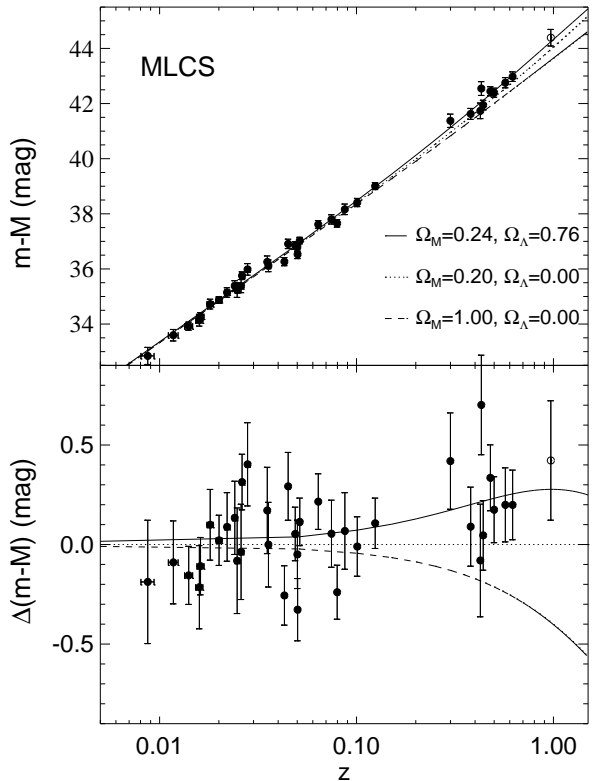}}
  \vskip1cm
  \caption{Hubble diagram (distance modulus vs.\ redshift)
  from the High-Z Supernova Team \cite{riess1}.  The lines represent
  predictions from the cosmological models with the specified 
  parameters.  The lower plot indicates the difference between
  observed distance modulus and that predicted in an open-universe
  model.}
  \label{riesshubble}
\end{figure}
\begin{figure}[t]
  \epsfxsize = 8cm
  \centerline{
  \epsfbox{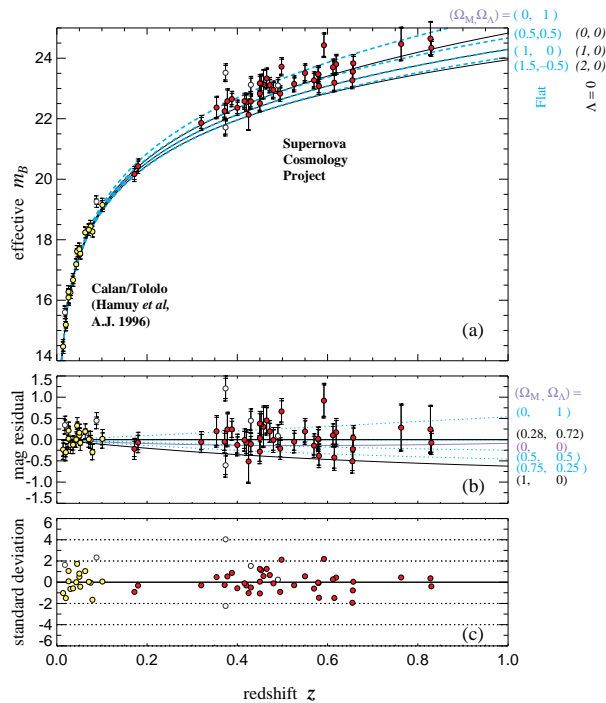}}
  \caption{Hubble diagram from the Supernova Cosmology Project
  \cite{perlmutter3}.  The bottom plot shows the number of standard
  deviations of each point from the best-fit curve.}
  \label{perlhubble}
\end{figure}

\begin{figure}[t]
  \epsfxsize = 7cm
  \centerline{
  \epsfbox{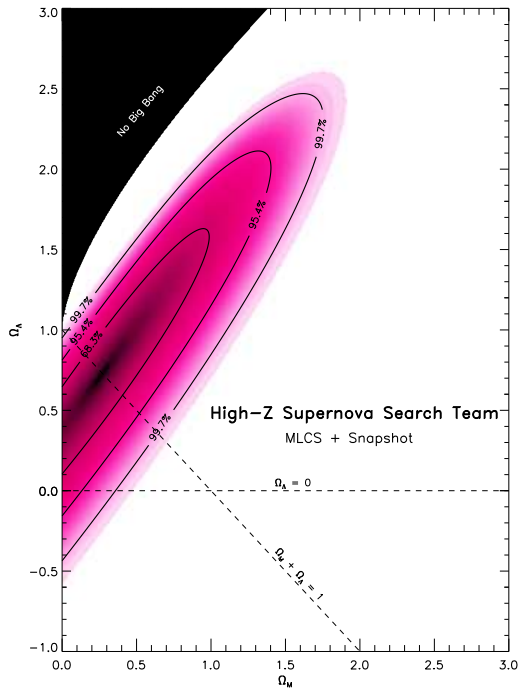}}
  \vskip.1cm
  \caption{Constraints in the $\Omega_{\rm M}$-$\Omega_\Lambda$
  plane from the High-Z Supernova Team
  \cite{riess1}.}
  \label{riessomegas}
\end{figure}
\begin{figure}[t]
  \epsfxsize = 7cm
  \centerline{
  \epsfbox{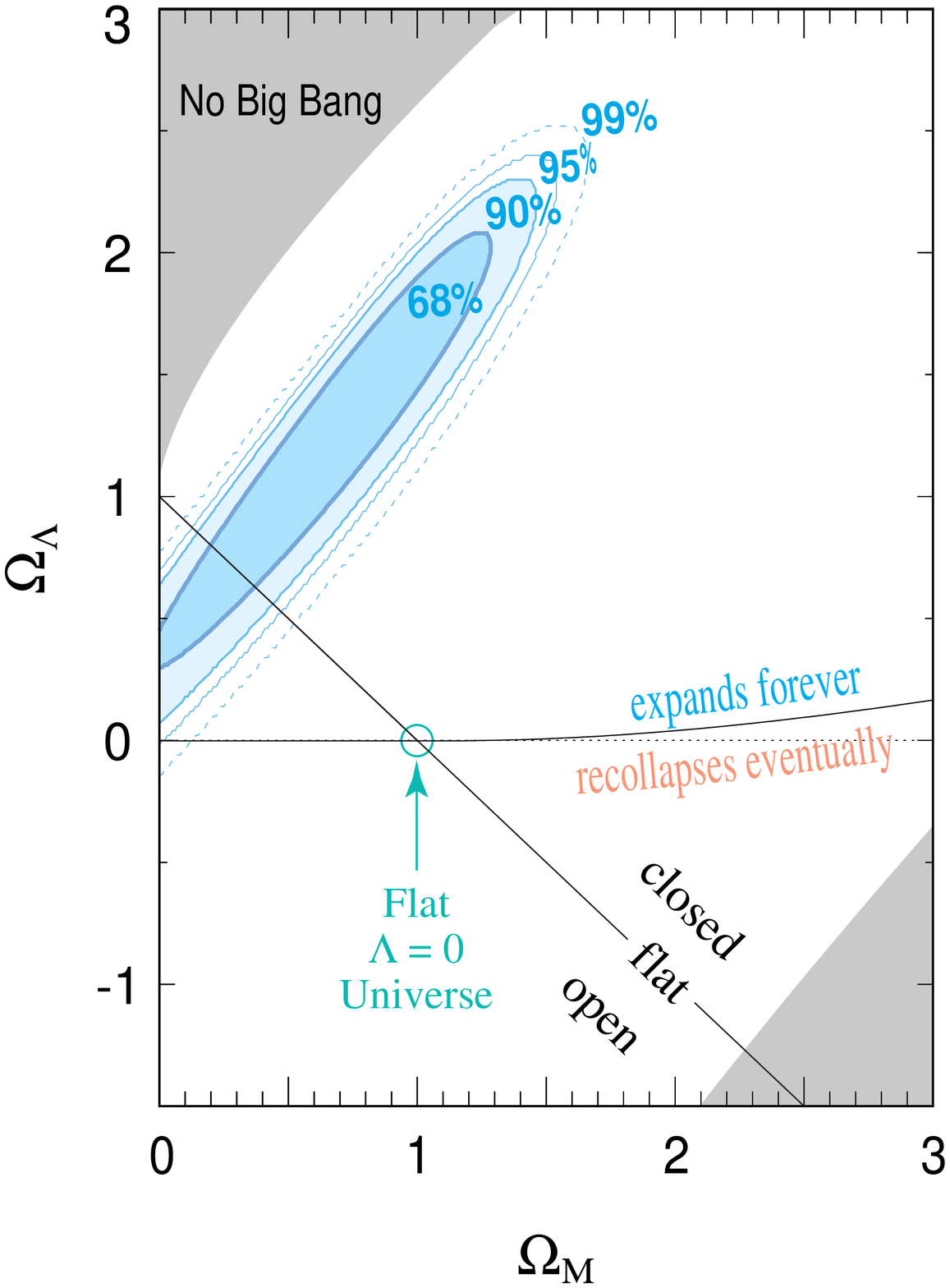}}
  \caption{Constraints in the $\Omega_{\rm M}$-$\Omega_\Lambda$
  plane from the Supernova Cosmology Project
  \cite{perlmutter3}.}
  \label{perlomegas}
\end{figure}

The fact that all SNe Ia are of similar intrinsic luminosities fits 
well with our understanding of these events as explosions which occur
when a white dwarf, onto which mass is gradually accreting from
a companion star, crosses the Chandrasekhar limit and explodes.
(It should be noted that our understanding of supernova
explosions is in a state of development, and theoretical models
are not yet able to accurately reproduce all of the important
features of the observed events.  See \cite{woosley,hachisu,hoflich}
for some recent work.)
The Chandrasekhar limit is a nearly-universal quantity, so it is
not a surprise that the resulting explosions are of nearly-constant
luminosity.  However, there is still a scatter of approximately
$40\% $ in the peak brightness observed in nearby supernovae, which
can presumably be traced to differences in the composition of the
white dwarf atmospheres.  Even if we could collect enough data
that statistical errors could be reduced to a minimum, the existence
of such an uncertainty would cast doubt on any attempts to study
cosmology using SNe~Ia as standard candles.  

Fortunately, the observed differences
in peak luminosities of SNe~Ia are very closely correlated with
observed differences in the shapes of their light curves:
dimmer SNe decline more rapidly after maximum brightness, while
brighter SNe decline more slowly \cite{phillips,rpk,hamuy}.
There is thus a one-parameter family of events, and measuring the
behavior of the light curve along with the apparent luminosity
allows us to largely correct for the intrinsic differences in
brightness, reducing the scatter from $40\% $ to less than $15\% $ 
--- sufficient precision to distinguish between cosmological models.
(It seems likely that the single parameter can be traced to the
amount of $^{56}$Ni produced in the supernova explosion; more
nickel implies both a higher peak luminosity and a higher 
temperature and thus opacity, leading to a slower decline.
It would be an exaggeration, however, to claim that this behavior
is well-understood theoretically.)

Following pioneering work reported in \cite{norgard},
two independent groups have undertaken searches for distant
supernovae in order to measure cosmological parameters.  Figure
(\ref{riesshubble}) shows the results for $m-M$ vs.\ $z$ for
the High-Z Supernova Team 
\cite{garnavich1,schmidt,riess1,garnavich2}, 
and Figure (\ref{perlhubble})
shows the equivalent results for the Supernova Cosmology Project
\cite{perlmutter1,perlmutter2,perlmutter3}.
Under the assumption that the energy density of the universe is
dominated by matter and vacuum components, these data can be
converted into limits on $\Omega_{\rm M}$ and $\Omega_\Lambda$,
as shown in Figures (\ref{riessomegas}) and (\ref{perlomegas}).

It is clear that the confidence intervals in the
$\Omega_{\rm M}$-$\Omega_\Lambda$ plane are consistent for the
two groups, with somewhat tighter constraints obtained by the 
Supernova Cosmology Project, who have more data points.
The surprising result is that both teams favor a positive
cosmological constant, and strongly rule out the traditional
$(\Omega_{\rm M}, \Omega_\Lambda) = (1,0)$ favorite universe.
They are even inconsistent with an open universe with zero
cosmological constant, given what we know about the matter
density of the universe (see below).

Given the significance of these results, it is natural to
ask what level of confidence we should have in them.
There are a number of potential sources of systematic
error which have been considered by the two teams; see
the original papers \cite{schmidt,riess1,perlmutter3}
for a thorough discussion.  The two most worrisome
possibilities are 
intrinsic differences between Type Ia supernovae at high
and low redshifts \cite{Drell:1999,Riess:1999}, and
possible extinction via intergalactic dust
\cite{aguirre1,aguirre2,aguirre3,sh,Totani:1999}.  (There is also the
fact that intervening weak lensing can change the 
distance-magnitude relation, but this seems to be a small
effect in realistic universes \cite{Holz:1998,Kantowski:1998}.)
Both effects have been carefully considered, and are thought
to be unimportant, although a better understanding will be
necessary to draw firm conclusions.  Here, I will briefly
mention some of the relevant issues.

As thermonuclear explosions of white dwarfs, Type Ia supernovae
can occur in a wide variety of environments.  Consequently, a
simple argument against evolution is that
the high-redshift environments, while chronologically younger,
should be a subset of all possible low-redshift environments,
which include regions that are ``young'' in terms of chemical
and stellar evolution.  Nevertheless, even a small amount of
evolution could ruin our ability to reliably constrain
cosmological parameters \cite{Drell:1999}.  In their original
papers \cite{schmidt,riess1,perlmutter3}, the supernova teams
found impressive consistency in the spectral and photometric
properties of Type Ia supernovae over a variety of redshifts
and environments ({\it e.g.}, in elliptical vs.\ spiral
galaxies).  More recently, however, Riess {\it et al.}\ 
\cite{Riess:1999} have presented tentative evidence for a
systematic difference in the properties of high- and low-redshift
supernovae, claiming that the risetimes (from initial explosion
to maximum brightness) were higher in the high-redshift
events.  Apart from the issue of whether the existing data support 
this finding, it is not immediately clear whether such a difference
is relevant to the distance determinations: first, because
the risetime is not used in 
determining the absolute luminosity at peak brightness, and
second, because a process which only affects the very early
stages of the light curve is most plausibly traced to differences
in the outer layers of the progenitor, which may have a 
negligible affect on the total energy output.  Nevertheless,
any indication of evolution could bring into question the fundamental
assumptions behind the entire program.  It is therefore essential
to improve the quality of both the data and the theories so that
these issues may be decisively settled.

Other than evolution, obscuration by dust is the leading concern
about the reliability of the supernova results.  Ordinary
astrophysical dust does not obscure equally at all wavelengths,
but scatters blue light preferentially, leading to the 
well-known phenomenon of ``reddening''.  Spectral measurements
by the two supernova teams reveal a negligible amount of reddening,
implying that any hypothetical dust must be a novel ``grey''
variety.  This possibility has been investigated by a number
of authors \cite{aguirre1,aguirre2,aguirre3,sh,Totani:1999}.
These studies have found that even grey dust is highly constrained
by observations: first, it is likely to be intergalactic rather
than within galaxies, or it would lead to additional dispersion
in the magnitudes of the supernovae; and second, intergalactic dust
would absorb ultraviolet/optical radiation and re-emit it at
far infrared wavelengths, leading to stringent constraints from
observations of the cosmological far-infrared background.  
Thus, while the possibility of obscuration has not been entirely
eliminated, it requires a novel kind of dust which is already
highly constrained (and may be convincingly ruled out by
further observations).

According to the best of our current understanding, then,
the supernova results indicating an accelerating universe seem
likely to be trustworthy.  Needless to say, however, the possibility
of a heretofore neglected systematic effect looms menacingly
over these studies.  Future experiments, including a proposed
satellite dedicated to supernova cosmology \cite{snap}, will both
help us improve our understanding of the physics of supernovae
and allow a determination of the distance/redshift relation to
sufficient precision to distinguish between the effects of a
cosmological constant and those of more mundane astrophysical
phenomena.  In the meantime, it is important to obtain independent
corroboration using other methods.

\subsection{Cosmic microwave background}
\label{section:cmb}

The discovery by the COBE satellite of temperature anisotropies
in the cosmic microwave background \cite{smoot} inaugurated a new
era in the determination of cosmological parameters.
To characterize the temperature fluctuations on the sky, we
may decompose them into spherical harmonics,
\begin{equation}
  {{\Delta T}\over T} = \sum_{lm} a_{lm} Y_{lm}(\theta,\phi)\ ,
\end{equation}
and express the amount of anisotropy at multipole moment $l$ via
the power spectrum,
\begin{equation}
  C_l = \langle |a_{lm}|^2 \rangle \ .
\end{equation}
Higher multipoles correspond to smaller angular separations
on the sky, $\theta = 180^\circ/l$.  
Within any given family of models, $C_l$ vs.\ $l$ will depend
on the parameters specifying the particular cosmology.
Although the case is far from closed, evidence has been mounting
in favor of a specific class of models --- those based on
Gaussian, adiabatic, nearly scale-free perturbations in a universe
composed of baryons, radiation, and cold dark matter.  (The
inflationary universe scenario \cite{Guth:1981,Linde:1982,Albrecht:1982}
typically predicts these kinds
of perturbations.)

\begin{figure}[t]
  \epsfxsize = 8cm
  \centerline{\epsfbox{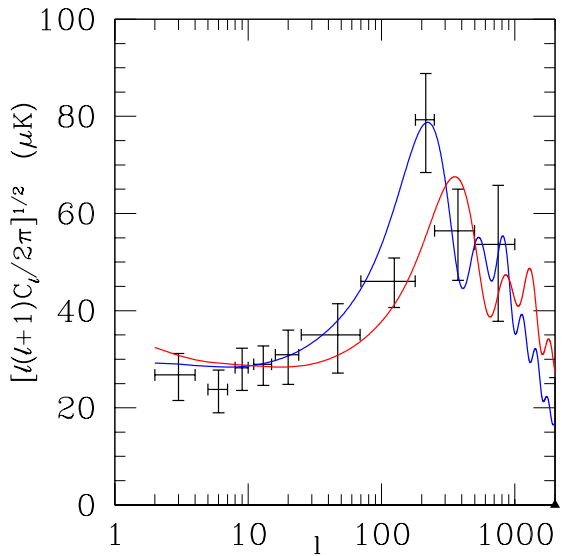}}
  \caption{CMB data (binned) and two theoretical curves:
  the model with a peak at $l\sim 200$ is a flat matter-dominated
  universe, while the one with a peak at $l\sim 400$ is an open
  matter-dominated universe.  From \cite{bjk}.}
  \label{knoxcmb}
\end{figure}

Although the dependence of the $C_l$'s on the parameters can be
intricate, nature has chosen not to test the patience of 
cosmologists, as one of the easiest features to measure --- the
location in $l$ of the first ``Doppler peak'', an increase in 
power due to acoustic oscillations --- provides one of the most
direct handles on the cosmic energy density, one of the most
interesting parameters.  The first
peak (the one at lowest $l$) corresponds to the angular scale
subtended by the Hubble radius $H_{\rm CMB}^{-1}$ at the time when the
CMB was formed (known variously as ``decoupling'' or ``recombination''
or ``last scattering'') \cite{Hu:1997qs}.
The angular scale at which we observe this peak is tied to the
geometry of the universe:  in a negatively (positively)
curved universe, photon paths diverge (converge), leading to
a larger (smaller) apparent angular size as compared to a
flat universe.  Since the scale $H_{\rm CMB}^{-1}$ is set
mostly by microphysics, this geometrical effect is dominant,
and we can relate the spatial curvature as characterized
by $\Omega$ to the observed peak in the CMB spectrum via
\cite{kss,Jungman:1996a,Hu:1996qz}
\begin{equation}
  l_{\rm peak} \sim 220\Omega^{-1/2}\ .
\end{equation}
More details about the spectrum (height of the peak, features
of the secondary peaks) will depend on other cosmological
quantities, such as the Hubble constant and the baryon density
\cite{Bond:1994,Hu:1995,Jungman:1996b,Zaldarriaga:1997}.

\begin{figure}[t]
  \epsfxsize = 8cm
  \centerline{\epsfbox{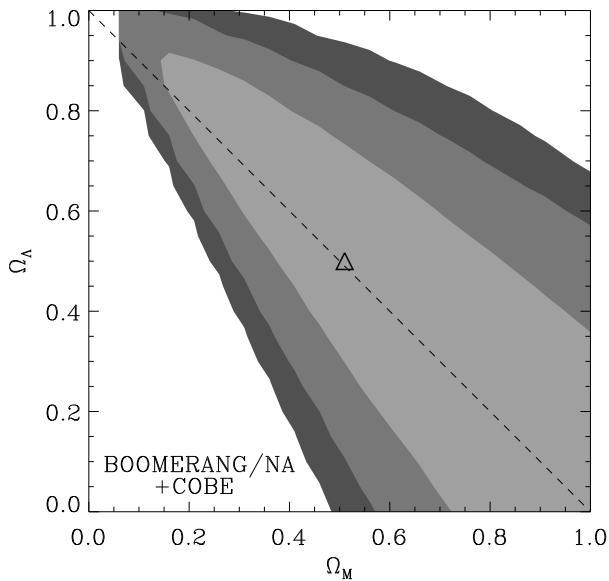}}
  \caption{Constraints in the $\Omega_{\rm M}$-$\Omega_\Lambda$
  plane from the North American flight of the BOOMERANG microwave 
  background balloon experiment.  
  From \cite{boom}.}
  \label{boomomegas}
\end{figure}

Figure \ref{knoxcmb} shows a summary of data as of 1998, with various
experimental results consolidated into bins, along with two
theoretical models.  Since that time, the data have continued to
accumulate (see for example \cite{Miller:1999,boom}), and the
near future should see a wealth of new results of ever-increasing
precision.  It is clear from the figure that there is good
evidence for a peak at approximately $l_{\rm peak} \sim 200$,
as predicted in a spatially-flat universe.  
This result can be made more quantitative by fitting the CMB
data to models with different values of $\Omega_{\rm M}$ and
$\Omega_\Lambda$ \cite{bjk,Bartlett:1998,Lineweaver:1998,
Ratra:1999,Dodelson:1999}, or by combining the CMB data with
other sources, such as supernovae or large-scale structure
\cite{White:1998,teh,garnavich2,Hu:1999,Tegmark:1999,Efstathiou:1998,
Bridle:1999,Bahcall:1999}.  Figure \ref{boomomegas}
shows the constraints from the CMB in the 
$\Omega_{\rm M}$-$\Omega_\Lambda$ plane, using data from the 1997 
test flight of the BOOMERANG experiment \cite{boom}. 
(Although the data used to make this plot are essentially independent
of those shown in the previous figure, the constraints obtained
are nearly the same.)  It is clear that the CMB data provide
constraints which are complementary to those obtained using
supernovae; the two approaches yield confidence contours which
are nearly orthogonal in the $\Omega_{\rm M}$-$\Omega_\Lambda$ 
plane.  The region of overlap is in the vicinity of
$(\Omega_{\rm M}, \Omega_\Lambda) = (0.3, 0.7)$, which we will
see below is also consistent with other determinations.

\subsection{Matter density}
\label{section:omegam}

Many cosmological tests, such as the two just discussed, will
constrain some combination of $\Omega_{\rm M}$ and $\Omega_\Lambda$.
It is therefore useful to consider tests of $\Omega_{\rm M}$ alone,
even if our primary goal is to determine $\Omega_\Lambda$.
(In truth, it is also hard to constrain $\Omega_{\rm M}$ alone,
as almost all methods actually constrain some combination
of $\Omega_{\rm M}$ and the Hubble constant $h = H_0
/(100$~km/sec/Mpc); the HST Key Project on the extragalactic
distance scale finds $h = 0.71 \pm 0.06$ \cite{mould},
which is consistent with other methods \cite{freedman}, and what
I will assume below.)

For years, determinations of $\Omega_{\rm M}$ based on dynamics
of galaxies and clusters have yielded values between approximately
$0.1$ and $0.4$ --- noticeably larger than the density parameter in
baryons as inferred from primordial nucleosynthesis, $\Omega_{\rm B}=
(0.019 \pm 0.001)h^{-2} \approx 0.04$ \cite{schramm,burles}, 
but noticeably smaller than the critical density.
The last several years have witnessed a number of new methods
being brought to bear on the question; the quantitative results
have remained unchanged, but our confidence in them has increased
greatly.

A thorough discussion of determinations of $\Omega_{\rm M}$
requires a review all its own, and good ones are available
\cite{Dekel:1996,Bahcall:1998,Turner:1999kz,freedman,prim99}.
Here I will just sketch some of the important methods.  

The traditional method to estimate
the mass density of the universe is to ``weigh'' a cluster of
galaxies, divide by its luminosity, and extrapolate the
result to the universe as a whole.  Although clusters are not
representative samples of the universe, they are sufficiently
large that such a procedure has a chance of working.  Studies
applying the virial theorem to cluster dynamics have typically
obtained values $\Omega_{\rm M} = 0.2 \pm 0.1$
\cite{carlberg96,Dekel:1996,Bahcall:1998}.  
Although it is possible that
the global value of $M/L$ differs appreciably from its value
in clusters, extrapolations from small scales do not seem
to reach the critical density \cite{Bahcall:1995}.  New
techniques to weigh the clusters, including gravitational
lensing of background galaxies \cite{smail} and temperature
profiles of the X-ray gas \cite{lewis}, while not yet in
perfect agreement with each other, reach essentially 
similar conclusions.

Rather than measuring the mass relative to the luminosity
density, which may be different inside and outside clusters,
we can also measure it with respect to the baryon density
\cite{white},
which is very likely to have the same value in clusters as
elsewhere in the universe, simply because there is no way
to segregate the baryons from the dark matter on such large
scales.  Most of the baryonic mass is in the hot intracluster
gas \cite{Fukugita:1997bi}, and the fraction $f_{\rm gas}$ 
of total mass in this form can be measured either by
direct observation of X-rays from the gas 
\cite{Mohr:1999} or by distortions
of the microwave background by scattering off hot electrons
(the Sunyaev-Zeldovich effect) \cite{carlstrom}, typically
yielding $0.1 \leq f_{\rm gas} \leq 0.2$.
Since primordial nucleosynthesis provides a determination
of $\Omega_{\rm B}\sim 0.04$, these measurements imply
\begin{equation}
  \Omega_{\rm M} = \Omega_{\rm B}/f_{\rm gas}
  = 0.3\pm 0.1 \ ,
\end{equation}
consistent with the value determined from mass to light ratios.

Another handle on the density parameter in matter comes from
properties of clusters at high redshift.  The very existence
of massive clusters has been used to argue in favor of 
$\Omega_{\rm M}\sim 0.2$ \cite{bf2}, and the lack of appreciable 
evolution of clusters from high redshifts to the present 
\cite{bfc,carlberg97} provides additional evidence that
$\Omega_{\rm M} < 1.0$.

The story of large-scale motions is more ambiguous.  The
peculiar velocities of galaxies are sensitive to the underlying
mass density, and thus to $\Omega_{\rm M}$, but also to the
``bias'' describing the relative amplitude of fluctuations in
galaxies and mass \cite{Dekel:1996,dekel97}.  Difficulties 
both in measuring the flows and in disentangling the mass density
from other effects make it difficult to draw conclusions at
this point, and at present it is hard to say much more
than $0.2 \leq \Omega_{\rm M} \leq 1.0$.

Finally, the matter density parameter can be extracted from
measurements of the power spectrum of density fluctuations
(see for example \cite{peacock}).  
As with the CMB, predicting
the power spectrum requires both an assumption of the correct
theory and a specification of a number of cosmological
parameters.  In simple models ({\it e.g.}, with only cold dark
matter and baryons, no massive neutrinos), the spectrum can be
fit (once the amplitude is normalized) by a single ``shape
parameter'', which is found to be equal to $\Gamma =
\Omega_{\rm M}h$.  (For more complicated models see
\cite{eh}.)  Observations then yield $\Gamma \sim 0.25$,
or $\Omega_{\rm M}\sim 0.36$.  For a more careful comparison
between models and observations, see 
\cite{Liddle:1993fq,Liddle:1996,dodelson,primack}.

Thus, we have a remarkable convergence on values for the
density parameter in matter:
\begin{equation}
  0.1 \leq \Omega_{\rm M} \leq 0.4\ .
\end{equation}
Even without the supernova results, this determination in
concert with the CMB measurements favoring a flat universe
provide a strong case for a nonzero cosmological constant.

\subsection{Gravitational lensing}
\label{section:lensing}

The volume of space back to a specified redshift, given by
(\ref{vol}), depends sensitively on $\Omega_\Lambda$.  Consequently,
counting the apparent density of observed objects, whose actual
density per cubic Mpc is assumed to be known, provides a potential
test for the cosmological constant \cite{gott,fukugita,turner90,cpt}.  
Like tests of distance vs.\ redshift, a significant problem for
such methods is the luminosity evolution of whatever objects
one might attempt to count.  A modern attempt to circumvent
this difficulty is to use the statistics of gravitational lensing
of distant galaxies; the hope is that the number of condensed
objects which can act as lenses is
less sensitive to evolution than the number of visible objects.

In a spatially flat universe,
the probability of a source at redshift $z_s$ being lensed, relative
to the fiducial ($\Omega_{\rm M}=1$, $\Omega_\Lambda = 0$) case,
is given by
\begin{equation}
  P_{\rm lens} = 
  {{15}\over 4}\left[1-(1+z_s)^{-1/2}\right]^{-3}
  \int_1^{a_s} {{H_0}\over{H(a)}} \left[{{d_A(0,a)d_A(a,a_s)}\over
  {d_A(0,a_s)}}\right]\, da \ ,
\end{equation}
where $a_s=1/(1+z_s)$.
\begin{figure}[t]
  \epsfxsize = 8cm
  \centerline{\epsfbox{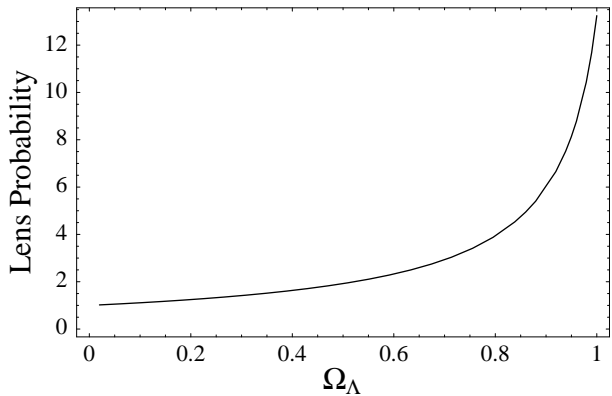}}
  \caption{Gravitational lens probabilities in a flat universe
  with $\Omega_{\rm M} + \Omega_\Lambda = 1$, relative to
  $\Omega_{\rm M} = 1$, $\Omega_\Lambda = 0$, for a source
  at $z=2$.}
  \label{glens2}
\end{figure}
As shown in Figure (\ref{glens2}), the probability rises dramatically
as $\Omega_\Lambda$ is increased to unity as we keep $\Omega$ fixed.
Thus, the absence of a large number of such lenses would imply an
upper limit on $\Omega_\Lambda$.

Analysis of lensing statistics is complicated by
uncertainties in evolution, extinction, and biases in the lens
discovery procedure.  It has been argued \cite{kochanek,falco} that
the existing data allow us to place an upper limit of 
$\Omega_\Lambda < 0.7$ in a flat 
universe.  However, other groups \cite{Chiba:1998,Cheng:1998} have 
claimed that the current data actually favor a nonzero cosmological
constant.  The near future will bring larger, more objective
surveys, which should allow these ambiguities to be resolved.
Other manifestations of lensing can also be used to constrain
$\Omega_\Lambda$, including statistics of giant arcs \cite{wu},
deep weak-lensing surveys \cite{hui}, and lensing in the Hubble
Deep Field \cite{cqm}.

\subsection{Other tests}
\label{section:othertests}

There is a tremendous variety of ways in which a nonzero cosmological
constant can manifest itself in observable phenomena.  Here is an
incomplete list of additional possibilities; see also 
\cite{cpt,cohn,sahni}.
\begin{itemize}
  \item Observations of numbers of objects vs.\ redshift are
  a potentially sensitive test of cosmological parameters
  if evolutionary effects can be brought under control.  Although
  it is hard to account for the luminosity evolution of 
  galaxies, it may be possible to indirectly count dark halos
  by taking into account the rotation speeds of visible
  galaxies, and upcoming redshift surveys could be used to
  constrain the volume/redshift relation \cite{newman}.
  \item Alcock and Paczynski \cite{alcock} showed that the 
  relationship between the apparent transverse and radial sizes 
  of an object of cosmological size depends on the 
  expansion history of the universe.  Clusters of galaxies would
  be possible candidates for such a measurement, but they are
  insufficiently isotropic; alternatives, however, have been
  proposed, using for example the quasar correlation function 
  as determined from redshift surveys
  \cite{phillips94,popowski}, or the Lyman-$\alpha$ forest
  \cite{hsb}.
  \item In a related effect, the dynamics of large-scale
  structure can be affected by a nonzero cosmological constant;
  if a protocluster, for example, is anisotropic, it can begin
  to contract along a minor axis while the universe is
  matter-dominated and along its major axis while the universe
  is vacuum-dominated.  Although small, such effects may be
  observable in individual clusters \cite{lahav} or in
  redshift surveys \cite{ballinger}.
  \item A different version of the distance-redshift test
  uses extended lobes of radio galaxies as modified
  standard yardsticks.  Current observations disfavor
  universes with $\Omega_{\rm M}$ near unity (\cite{guerra},
  and references therein).
  \item Inspiralling compact binaries at cosmological distances 
  are potential sources of gravitational waves.  It turns out
  that the redshift distribution of events is sensitive to
  the cosmological constant; although speculative, it has been
  proposed that advanced LIGO detectors could use this effect
  to provide measurements of $\Omega_\Lambda$ \cite{Wang:1997}.
  \item Finally, consistency of the age of the universe and the
  ages of its oldest constituents is a classic test of the
  expansion history.  If stars were sufficiently old and 
  $H_0$ and $\Omega_{\rm M}$ were sufficiently high, a positive
  $\Omega_\Lambda$ would be necessary to reconcile the two,
  and this situation has occasionally been thought to hold.
  Measurements of geometric parallax to nearby stars from the 
  Hipparcos satellite have, at the least, called into question
  previous determinations of the ages of the oldest globular clusters,
  which are now thought to be perhaps 12 billion rather than
  15 billion years old (see the discussion in \cite{freedman}).
  It is therefore unclear whether the age issue forces a 
  cosmological constant upon us, but by now it seems forced
  upon us for other reasons.
\end{itemize}

\section{Physics issues}
\label{section:physics}

In Section (\ref{section:vacuumenergy}) we discussed the large 
difference between the magnitude of the 
vacuum energy expected from zero-point
fluctuations and scalar potentials, $\rho_{\Lambda}^{\rm theor}
\sim 2\times 10^{110} {\rm ~erg/cm}^3$,
and the value we apparently observe,
$\rho_\Lambda^{\rm (obs)}\sim 2\times 10^{-10} {\rm ~erg/cm}^3$
(which may be thought of as an upper limit, if we wish
to be careful).  It is somewhat unfair to characterize this
discrepancy as a factor of $10^{120}$, since energy density
can be expressed as a mass scale to the fourth power.
Writing $\rho_{\Lambda} = M_{\rm vac}^4$,
we find $M_{\rm vac}^{\rm (theory)}\sim M_{\rm Pl}
\sim 10^{18}$~GeV and 
$M_{\rm vac}^{\rm (obs)} \sim 10^{-3}$~eV, so a more fair
characterization of the problem would be
\begin{equation}
  {M_{\rm vac}^{\rm (theory)} \over M_{\rm vac}^{\rm (obs)}}
  \sim 10^{30}\ .  
  \label{naive}
\end{equation}
Of course, thirty orders of magnitude still constitutes a
difference worthy of our attention.

Although the mechanism which suppresses the naive value
of the vacuum energy is unknown, it seems easier to imagine
a hypothetical scenario which makes it exactly zero than
one which sets it to just the right value to be observable
today.  (Keeping in mind that it is the zero-temperature,
late-time
vacuum energy which we want to be small; it is expected to
change at phase transitions, and a large value in the early
universe is a necessary component of inflationary universe
scenarios \cite{Guth:1981,Linde:1982,Albrecht:1982}.)
If the recent observations pointing toward a cosmological constant
of astrophysically relevant magnitude are confirmed, we will
be faced with the challenge of explaining not only why the
vacuum energy is smaller than expected, but also why it has
the specific nonzero value it does.

\subsection{Supersymmetry}
\label{section:susy}

Although initially investigated for other reasons, supersymmetry
(SUSY) turns out to have a significant impact on the cosmological
constant problem, and may even be said to solve it halfway.
SUSY is a spacetime symmetry relating fermions and bosons to each
other.  Just as ordinary symmetries are associated with
conserved charges, supersymmetry is associated with
``supercharges'' $Q_\alpha$, where $\alpha$
is a spinor index (for introductions see \cite{nilles,lykken,martin}).  
As with ordinary symmetries, a theory may
be supersymmetric even though a given state is not supersymmetric;
a state which is annihilated by the supercharges,
$Q_\alpha |\psi\rangle =0$, preserves supersymmetry, while states
with $Q_\alpha |\psi\rangle \neq 0$ are said to spontaneously
break SUSY.

Let's begin by considering ``globally supersymmetric'' theories,
which are defined in flat spacetime (obviously an
inadequate setting in which to discuss the cosmological 
constant, but we have to start somewhere).  Unlike most
non-gravitational field theories, in supersymmetry 
the total energy of a state has an absolute meaning; the
Hamiltonian is related to the supercharges in a straightforward
way:
\begin{equation}
  H = \sum_\alpha \{ Q_\alpha, Q_\alpha^\dagger \} \ ,
\end{equation}
where braces represent the anticommutator.  Thus, in a completely
supersymmetric state (in which $Q_\alpha |\psi\rangle =0$ for all
$\alpha$), the energy vanishes automatically,
$\langle \psi |H| \psi\rangle=0$ \cite{Zumino:1975}.  
More concretely, in a given
supersymmetric theory we can explicitly calculate the contributions
to the energy from vacuum fluctuations and from the scalar potential
$V$.  In the case of vacuum fluctuations, contributions
from bosons are exactly canceled by equal and opposite contributions 
from fermions when supersymmetry is unbroken.  Meanwhile,
the scalar-field potential in supersymmetric theories takes
on a special form; scalar fields $\phi^i$ must be complex
(to match the degrees of freedom of the fermions), and
the potential is derived from a function called the superpotential
$W(\phi^i)$ which is necessarily holomorphic (written in terms
of $\phi^i$ and not  its complex
conjugate $\bar\phi^{i}$).  In the simple Wess-Zumino
models of spin-0 and spin-1/2 fields, for example, the scalar
potential is given by
\begin{equation}
  V(\phi^i, \bar\phi^j) = \sum_i |\partial_i W|^2\ ,
  \label{susypotl}
\end{equation}
where $\partial_iW = \partial W/\partial \phi^i$.  In such a 
theory, one can show that SUSY will be unbroken only for values
of $\phi^i$ such that $\partial_i W=0$, implying 
$V(\phi^i, \bar\phi^j) =0$.

So the vacuum energy of a supersymmetric state in a globally
supersymmetric theory will vanish.  This represents rather less
progress than it might appear at first sight, since: 1.) Supersymmetric
states manifest a degeneracy in the mass spectrum of bosons and
fermions, a feature not apparent in the observed world; and 2.)
The above results imply that non-supersymmetric states have a 
positive-definite vacuum energy.  Indeed, in a state where
SUSY was broken at an energy scale $M_{\rm SUSY}$, we would
expect a corresponding vacuum energy $\rho_{\Lambda}\sim
M_{\rm SUSY}^4$.  In the real world, the fact that 
accelerator experiments have not
discovered superpartners for the known particles of the Standard
Model implies that 
$M_{\rm SUSY}$ is of order $10^3$~GeV or higher.  Thus, we
are left with a discrepancy
\begin{equation}
  {{M_{\rm SUSY}}\over{M_{\rm vac}}} \geq 10^{15}\ .
\end{equation}
Comparison of this discrepancy with the naive discrepancy 
(\ref{naive})
is the source of the claim that SUSY can solve the cosmological
constant problem halfway (at least on a log scale).

As mentioned, however, this analysis is strictly valid only
in flat space.  In curved spacetime, the global transformations
of ordinary supersymmetry are promoted to the position-dependent 
(gauge) transformations of supergravity.  In this context the 
Hamiltonian and supersymmetry generators play different
roles than in flat spacetime, but it is still possible to
express the vacuum energy in terms of a scalar field potential
$V(\phi^i, \bar\phi^{j})$.  In supergravity $V$ depends
not only on the superpotential $W(\phi^i)$, but also on a
``K\"ahler potential'' $K(\phi^i, \bar\phi^j)$, and the 
K\"ahler metric $K_{i{\bar\jmath}}$ constructed from the 
K\"ahler potential by $K_{i{\bar\jmath}} = \partial^2 K
/\partial\phi^i \partial\bar\phi^j$.  (The basic role
of the K\"ahler metric is to define the kinetic term
for the scalars, which takes the form $g^{\mu\nu}
K_{i{\bar\jmath}}\partial_\mu\phi^i \partial_\nu\bar\phi^j$.)
The scalar potential is
\begin{equation}
  V(\phi^i, \bar\phi^j) = e^{K/M_{\rm Pl}^2}
  \left[K^{i{\bar\jmath}}(D_iW) (D_{\bar\jmath} \bar{W}) 
  - 3M_{\rm Pl}^{-2} |W|^2\right] \ ,
\end{equation}
where $D_iW$ is the K\"ahler derivative,
\begin{equation}
  D_i W = \partial_i W + M_{\rm Pl}^{-2} (\partial_i K) W\ .
\end{equation}
Note that, if we take the canonical K\"ahler metric
$K_{i{\bar\jmath}}=\delta_{i{\bar\jmath}}$, in the limit $M_{\rm Pl}
\rightarrow \infty$ ($G\rightarrow 0$) the first term in
square brackets reduces to the flat-space
result (\ref{susypotl}).  But with gravity, in addition to the
non-negative first term we find a second term
providing a non-positive contribution.  Supersymmetry is
unbroken when  $D_iW=0$; the effective cosmological constant
is thus non-positive.  We are therefore free to imagine a
scenario in which supersymmetry is broken in exactly the
right way, such that the two terms in parentheses cancel to
fantastic accuracy, but only at the cost of an
unexplained fine-tuning (see for example
\cite{Cremmer:1983}).  At the same time, supergravity is
not by itself a renormalizable quantum theory, and therefore
it may not be reasonable to hope that a solution can be
found purely within this context.

\subsection{String theory}
\label{section:strings}

Unlike supergravity, string theory appears to be a 
consistent and well-defined theory of quantum gravity, and
therefore calculating the value of the cosmological constant
should, at least in principle, be possible.  On the
other hand, the number of vacuum states seems to be quite large,
and none of them (to the best of our current knowledge) features
three large spatial dimensions, broken supersymmetry, and a small
cosmological constant.  At the same time, there are reasons to
believe that any realistic vacuum of string theory must be
strongly coupled \cite{dineseiberg}; therefore, our inability
to find an appropriate solution may simply be due to the
technical difficulty of the problem.  
(For general introductions to string theory, see
\cite{gsw,polchinski}; for cosmological issues, see 
\cite{Lykken:1998,Banks:1999}).

String theory is naturally formulated in more than four spacetime
dimensions.  Studies of duality symmetries have revealed that 
what used to be thought of as five distinct ten-dimensional
superstring theories --- Type I, Types IIA and IIB, and heterotic
theories based on gauge groups E(8)$\times$E(8) and SO(32) ---
are, along with eleven-dimensional supergravity, different
low-energy weak-coupling limits of a single underlying theory,
sometimes known as M-theory.  In each of these six cases, the
solution with the maximum number of uncompactified, flat spacetime
dimensions is a stable vacuum preserving all of the supersymmetry.
To bring the theory closer to the world we observe, 
the extra dimensions
can be compactified on a manifold whose Ricci tensor vanishes.
There are a large number of possible compactifications, many of
which preserve some but not all of the original supersymmetry.
If enough SUSY is preserved, the vacuum energy will remain zero;
generically there will be a manifold of such states, known as
the moduli space.

Of course, to describe our world we want to break all of the
supersymmetry.  Investigations in contexts where this can be
done in a controlled way have found that the induced cosmological
constant vanishes at the classical level, but a substantial
vacuum energy is typically induced by quantum corrections \cite{gsw}.
Moore \cite{Moore:1987} has suggested that Atkin-Lehner symmetry, which
relates strong and weak coupling on the string
worldsheet, can enforce the vanishing
of the one-loop quantum contribution in certain models
(see also \cite{Dienes:1990a,Dienes:1990b});
generically, however, there would still be an appreciable
contribution at two loops.

Thus, the search is still on for a four-dimensional string theory
vacuum with broken supersymmetry and vanishing (or very small)
cosmological constant.  (See \cite{Dine:1999} for a general
discussion of the vacuum problem in string theory.)
The difficulty of achieving this in
conventional models has inspired a number of more speculative
proposals, which I briefly list here.
\begin{itemize}
 \item In three spacetime dimensions supersymmetry can
  remain unbroken, maintaining a zero cosmological constant,
  in such a way as to break the mass degeneracy between bosons
  and fermions \cite{Witten:1995a}.  This mechanism relies crucially 
  on special properties of spacetime in (2+1) dimensions, but in
  string theory it sometimes happens that the strong-coupling
  limit of one theory is another theory in one higher dimension
  \cite{Witten:1995b,Witten:1995c}.  
 \item More generally, it is now understood that (at least in
  some circumstances) string theory obeys the ``holographic
  principle'', the idea that a theory with gravity in $D$ 
  dimensions is equivalent to a theory without gravity in
  $D-1$ dimensions \cite{'tHooft:1993,Susskind:1995}.  
  In a holographic theory, the number of
  degrees of freedom in a region grows as the area of its
  boundary, rather than as its volume.  Therefore, the 
  conventional computation of the cosmological constant due to
  vacuum fluctuations conceivably involves a vast overcounting
  of degrees of freedom.  We might imagine that a more correct
  counting would yield a much smaller estimate of the vacuum
  energy \cite{Banks:1995,Cohen:1999,Verlinde:1999,schmidhuber}, 
  although no 
  reliable calculation has been done as yet.
 \item The absence of manifest SUSY in our world leads us to ask
  whether the beneficial aspect of canceling contributions to the
  vacuum energy could be achieved even without a truly supersymmetric
  theory.  Kachru, Kumar and Silverstein \cite{Kachru:1999} have
  constructed such a string theory, and argue that the 
  perturbative contributions to the cosmological constant should
  vanish (although the actual calculations are somewhat delicate,
  and not everyone agrees \cite{Iengo:1999}).
  If such a model could be made to work, it is possible that
  small non-perturbative effects could generate a cosmological
  constant of an astrophysically plausible magnitude
  \cite{harvey}.
 \item A novel approach to compactification starts by imagining
  that the fields of the Standard Model are confined to a 
  (3+1)-dimensional manifold (or ``brane'', in string theory
  parlance) embedded in a larger space.  While gravity is
  harder to confine to a brane, phenomenologically acceptable
  scenarios can be constructed if either the extra dimensions
  are any size less than a millimeter
  \cite{rubakov,antoniadis,Horava:1996,add,Kakushadze:1999},
  or if there is significant spacetime curvature in a 
  non-compact extra dimension  \cite{visser,rs,gog}.
  Although these scenarios do not offer a simple solution to
  the cosmological constant problem, the relationship between
  the vacuum energy and the expansion rate can differ from our
  conventional expectation (see for example
  \cite{Binetruy:1999,Kanti:1999}), 
  and one is free to imagine that 
  further study may lead to a solution in this context
  (see for example \cite{Steinhardt:1999eh,Burgess:1999}).
\end{itemize}

Of course, string theory might not be the correct description of
nature, or its current formulation might not be directly relevant
to the cosmological constant problem.  For example, a solution
may be provided by loop quantum gravity \cite{Gambini:1998},
or by a composite graviton \cite{Sundrum:1999}.
It is probably safe to believe that a significant advance in
our understanding of fundamental physics will be required before
we can demonstrate the existence of a vacuum state with the
desired properties.  (Not to mention the equally important
question of why our world is based on such a state, rather than
one of the highly supersymmetric states that appear to be perfectly
good vacua of string theory.)

\subsection{The anthropic principle}
\label{section:anthropic}

The anthropic principle \cite{barrow,Hogan:1999} is essentially
the idea that some of the parameters characterizing the universe
we observe may not be determined directly by the fundamental
laws of physics, but also by the truism that intelligent
observers will only ever experience conditions which allow
for the existence of intelligent observers.  Many professional
cosmologists view this principle in much the same way as
many traditional literary critics view deconstruction --- as
somehow simultaneously empty of content and capable of 
working great evil.  Anthropic arguments are easy to misuse,
and can be invoked as a way out of doing the hard work
of understanding the real reasons behind why we observe the
universe we do.  Furthermore, a sense of disappointment
would inevitably accompany the realization that there were
limits to our ability to unambiguously and directly
explain the observed universe from first principles.  
It is nevertheless possible that some features of our world
have at best an anthropic explanation, and the value of the
cosmological constant is perhaps the most likely candidate.

In order for the tautology that ``observers will only observe
conditions which allow for observers'' to have any force,
it is necessary for there to be alternative conditions ---
parts of the universe, either in space, time, or branches of the
wavefunction --- where things are different.  In such a case, our
local conditions arise as some combination of the relative
abundance of different environments and the likelihood that
such environments would give rise to intelligence.  Clearly,
the current state of the art doesn't allow us to characterize
the full set of conditions in the entire universe with any
confidence, but modern theories of inflation and quantum
cosmology do at least allow for the possibility of widely
disparate parts of the universe in which the ``constants of
nature'' take on very different values (for recent examples see
\cite{Linde:1995ck,Vilenkin:1995yd,Linde:1994xx,
Hawking:1998,Linde:1998gs,Turok:1998he,Vilenkin:1998}).
We are therefore faced with the task of estimating quantitatively
the likelihood of observing any specific value of $\Lambda$
within such a scenario.

The most straightforward anthropic constraint on the vacuum
energy is that it must not be so high that galaxies never
form \cite{Weinberg:1987}.  From the discussion in Section
(\ref{section:structure}), we know that overdense regions do not 
collapse once the cosmological constant begins to dominate
the universe; if this happens before the epoch of galaxy formation,
the universe will be devoid of galaxies, and thus of stars and
planets, and thus (presumably) of intelligent life.  The 
condition that $\Omega_\Lambda(z_{\rm gal}) \leq
\Omega_{\rm M}(z_{\rm gal})$ implies
\begin{equation}
  {{\Omega_{\Lambda 0}}\over \Omega_{{\rm M}0}}
  \leq a_{\rm gal}^{-3} = (1+z_{\rm gal})^3 \sim 125\ ,
\end{equation}
where we have taken the redshift of formation of the first
galaxies to be $z_{\rm gal}\sim 4$.  Thus, the cosmological
constant could be somewhat larger than observation allows and
still be consistent with the existence of galaxies.  (This
estimate, like the ones below, holds parameters such as the
amplitude of density fluctuations fixed while allowing
$\Omega_\Lambda$ to vary; depending on one's model of the
universe of possibilities, it may be more defensible to vary
a number of parameters at once.  See for example 
\cite{Tegmark:1998,Garriga:2000,Hogan:1999}.)

However, it is better to ask what is most likely value of
$\Omega_\Lambda$, {\it i.e.} what is the value that would
be experienced by the largest number of observers 
\cite{Vilenkin:1995,Efstathiou:1995}?  Since a universe with
$\Omega_{\Lambda 0}/\Omega_{{\rm M}0}\sim 1$ will have many
more galaxies than one with $\Omega_{\Lambda 0}/\Omega_{{\rm M}0}
\sim 100$, it is quite conceivable that most observers will
measure something close to the former value.  The probability
measure for observing a value of $\rho_\Lambda$ can be
decomposed as
\begin{equation}
  d{\cal P}(\rho_\Lambda) = \nu(\rho_\Lambda)
  {\cal P}_*(\rho_\Lambda) d\rho_\Lambda\ ,
\end{equation}
where ${\cal P}_*(\rho_\Lambda) d\rho_\Lambda$ is
the {\it a priori} probability measure (whatever that might
mean) for $\rho_\Lambda$, and $\nu(\rho_\Lambda)$
is the average number of galaxies which form at the specified
value of $\rho_\Lambda$.  Martel, Shapiro and Weinberg
\cite{Martel:1997} have presented a calculation of 
$\nu(\rho_\Lambda)$ using a spherical-collapse model.
They argue that it is natural to take the {\it a priori} 
distribution to
be a constant, since the allowed range of $\rho_\Lambda$
is very far from what we would expect from particle-physics
scales.  Garriga and Vilenkin \cite{Garriga:1999} argue on
the basis of quantum cosmology that there can be a significant
departure from a constant {\it a priori} distribution.  However,
in either case the conclusion is that an observed 
$\Omega_{\Lambda 0}$ of the same order of magnitude as
$\Omega_{{\rm M}0}$ is by no means extremely unlikely (which is
probably the best one can hope to say given the uncertainties
in the calculation).

Thus, if one is willing to make the leap of
faith required to believe that the value of the cosmological
constant is chosen from an ensemble of possibilities, it is
possible to find an ``explanation'' for its current value (which,
given its unnaturalness from a variety of perspectives, seems
otherwise hard to understand).  Perhaps the most significant 
weakness of this point of view is the assumption that there
are a continuum of possibilities for the vacuum energy density.
Such possibilities correspond to choices of vacuum states
with arbitrarily similar energies.  If these states were
connected to each other, there would be local fluctuations which
would appear to us as massless fields, which are not observed
(see Section \ref{section:darkenergy}).  If on the other hand the vacua 
are disconnected, it is hard to understand why all possible values
of the vacuum energy are represented, rather than the differences
in energies between different vacua being given by some 
characteristic particle-physics scale such as 
$M_{\rm Pl}$ or $M_{\rm SUSY}$.  (For one scenario featuring
discrete vacua with densely spaced energies, see \cite{Banks:1991mb}.)
It will therefore (again) require advances in our understanding
of fundamental physics before an anthropic explanation for the
current value of the cosmological constant
can be accepted.

\subsection{Miscellaneous adjustment mechanisms}
\label{section:misc}

The importance of the cosmological constant problem has engendered 
a wide variety of proposed solutions.
This section will present only a brief outline of some of
the possibilities, along with references to recent
work; further discussion and references can be found in
\cite{weinberg,cpt,sahni}.

One approach which has received a great deal
of attention is the famous suggestion by
Coleman \cite{Coleman:1988}, that effects of virtual wormholes
could set the cosmological constant to zero at low energies.
The essential idea is that wormholes (thin tubes of spacetime
connecting macroscopically large regions) can act to change
the effective value of all the observed constants of nature.
If we calculate the wave function of the universe by performing
a Feynman path integral over all possible spacetime
metrics with wormholes, the dominant contribution will be from
those configurations whose effective values for the physical
constants extremize the action.  These turn out to be, under
a certain set of assumed properties of Euclidean 
quantum gravity, configurations with zero cosmological constant
at late times.  Thus, quantum cosmology predicts that the
constants we observe are overwhelmingly likely to take on values
which imply a vanishing total vacuum energy.
However, subsequent investigations have failed to inspire
confidence that the desired properties of Euclidean quantum
cosmology are likely to hold, although it is still something
of an open question; see discussions in \cite{weinberg,cpt}.

Another route one can take is to consider alterations of
the classical theory of gravity.  The simplest possibility
is to consider adding a scalar field to the theory, with
dynamics which cause the scalar to evolve to a value for
which the net cosmological constant vanishes (see for example 
\cite{Dolgov:1985,Starobinsky:1998}).  Weinberg, however, has pointed
out on fairly general grounds that such attempts are unlikely
to work \cite{weinberg,Weinberg:1996}; in models proposed
to date, either there is no solution for which the effective
vacuum energy vanishes, or there is a solution but with other
undesirable properties (such as making Newton's constant $G$
also vanish).  Rather than adding scalar fields, a related
approach is to remove degrees of freedom by making the
determinant of the metric, which multiplies $\Lambda_0$ in
the action (\ref{action}), a non-dynamical quantity, or at
least changing its dynamics in some way
(see \cite{Guendelman:1997,Wilczek:1998,Ng:1999}
for recent examples).  While this approach has not led to
a believable solution to the cosmological constant problem, it
does change the context in which it appears, and may induce
different values for the effective vacuum energy in different
branches of the wavefunction of the universe.

Along with global supersymmetry, there is one other symmetry
which would work to prohibit a cosmological constant: conformal
(or scale) invariance, under which the metric is multiplied
by a spacetime-dependent function, $g_{\mu\nu} \rightarrow
e^{\lambda(x)} g_{\mu\nu}$.  Like supersymmetry, conformal
invariance is not manifest in the Standard Model of particle
physics.  However, it has been proposed that quantum effects
could restore conformal invariance on length scales comparable
to the cosmological horizon size,
working to cancel the cosmological constant (for some examples see 
\cite{Tomboulis:1990gw,Antoniadis:1992fa,Antoniadis:1998fi}).
At this point it remains unclear whether this suggestion
is compatible with a more complete understanding of quantum gravity,
or with standard cosmological observations.

A final mechanism to suppress the cosmological constant,
related to the previous one,
relies on quantum particle production in de~Sitter space
(analogous to Hawking radiation around black holes).  
The idea is that the effective energy-momentum tensor of 
such particles may act to cancel out the bare cosmological
constant (for recent attempts see 
\cite{Tsamis:1993,Tsamis:1996,Abramo:1997,Ozer:1998}).
There is currently no consensus on whether such an
effect is physically observable (see for example
\cite{Unruh:1998}).

If inventing a theory in which the vacuum energy vanishes is 
difficult, finding a model that predicts a vacuum energy which is 
small but not quite zero is all that much harder.  Along these lines,
there are various numerological games one can play.  For example,
the fact that supersymmetry solves the problem halfway could be
suggestive; a theory in which the effective vacuum energy scale
was given not by $M_{\rm SUSY}\sim 10^3$~GeV but by 
$M_{\rm SUSY}^2/M_{\rm Pl}\sim 10^{-3}$~eV would seem to fit
the observations very well.  The challenging part of this
program, of course, is to 
devise such a theory.  Alternatively, one could imagine that
we live in a ``false vacuum'' --- that the absolute minimum of the
vacuum energy is truly zero, but we live in a state which is
only a local minimum of the energy.  Scenarios along these
lines have been explored 
\cite{Turner:1982,Garretson:1993,Kusenko:1997}; the major
hurdle to be overcome is explaining why the energy difference
between the true and false vacua is so much smaller than one
would expect.

\subsection{Other sources of dark energy}
\label{section:darkenergy}

Although a cosmological constant is an excellent fit to the
current data, the observations can also be accommodated by
any form of ``dark energy'' which does not cluster
on small scales (so as to avoid being detected by measurements
of $\Omega_{\rm M}$) and redshifts away only very slowly as the
universe expands [to account for the accelerated expansion,
as per equation (\ref{q})].  This possibility has been
extensively explored of late, and a number of candidates
have been put forward.

One way to parameterize such a component $X$ is by an effective
equation of state, $p_X = w_X \rho_X$.  
(A large number of phenomenological models of this type have
been investigated, starting with the early work in
\cite{Ozer:1987,Freese:1987}; see \cite{Overduin:1998,sahni} for 
many more references.)  The relevant range for
$w_X$ is between $0$ (ordinary matter) and $-1$ (true cosmological
constant); sources with $w_X > 0$ redshift away more rapidly 
than ordinary matter (and therefore cause extra deceleration),
while $w_X < -1$ is unphysical by the criteria discussed in
Section \ref{section:parameters} (although see
\cite{Caldwell:1999}).  While not every source
will obey an equation of state with $w_X=$~constant, it is
often the case that a single effective $w_X$ characterizes
the behavior for the redshift range over which the component
can potentially be observed.  
\begin{figure}[t]
  \epsfxsize = 13cm
  \centerline{\epsfbox{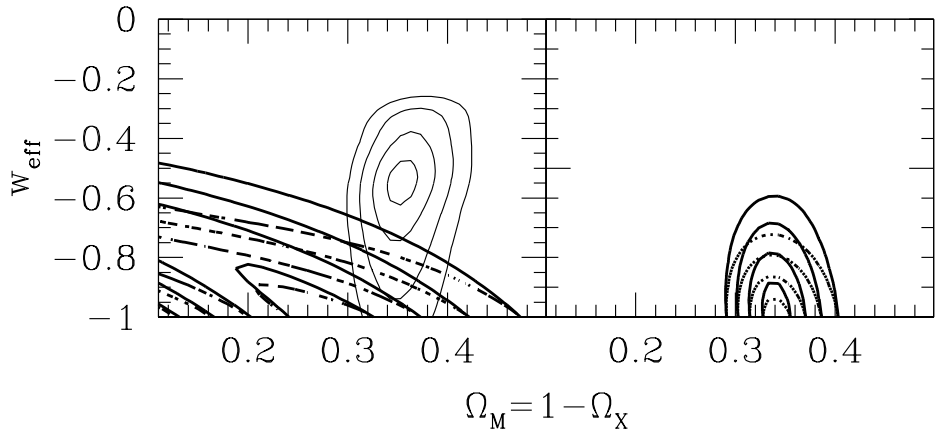}}
  \caption{Limits from supernovae and large-scale structure
  data on $\Omega_{\rm M}$ and the equation-of-state parameter
  $w_X$, in a flat universe dominated by matter and dark energy.
  Thin contours (on the left) represent limits from CMB and
  large-scale structure measurements, while thick contours are
  those from SNe observations; solid lines apply to models with
  constant $w_X$, while dashed lines apply to models of
  dynamical scalar fields.  The constraints are portrayed
  separately on the left, and combined on the right.
  From \cite{Perlmutter:1999}.}
  \label{wlimits}
\end{figure}
Current observations of supernovae, large-scale structure, 
gravitational lensing, and the CMB already
provide interesting limits on $w_X$ 
\cite{Ratra:1992,Coble:1997,tw97,Frieman:1998,
CSN98,garnavich2,perlmutter3,Waga:1999,
Perlmutter:1999,Wang:1999,Efstathiou:1999,Podariu}, 
and future data will be able to do much better
\cite{Efstathiou:1999,Huterer:1999,Cooray:1999,Saini:1999ba}.
Figure~(\ref{wlimits}) shows an example, in this case limits
from supernovae and large-scale structure 
on $w_X$ and $\Omega_{\rm M}$ in a universe which is
assumed to be flat and dominated by $X$ and ordinary
matter.  It is clear that the favored value for the 
equation-of-state parameter is near $-1$, that of a true cosmological
constant, although other values are not completely ruled out.

The simplest physical model for an appropriate dark
energy component is a single slowly-rolling scalar field,
sometimes referred to as ``quintessence''
\cite{Dolgov:1982gh,Weiss:1987,Peebles:1988,Ratra:1988,Wetterich:1988,
Hill:1989,Fujii:1990,
Frieman:1992,Frieman:1995,Ferreira:1998,Caldwell:1998,Huey:1999}.
In an expanding universe, a spatially homogeneous scalar with potential 
$V(\phi)$ and minimal coupling to gravity obeys
\begin{equation}
  \ddot\phi + 3 H \dot\phi + V'(\phi) = 0\ ,
\end{equation}
where $H$ is the Hubble parameter, overdots indicate time
derivatives, and primes indicate derivatives with respect to
$\phi$.  This equation is similar to (\ref{deltaeq}),
with analogous solutions.  The Hubble parameter acts as a
friction term; for generic potentials, the field will be 
overdamped (and thus approximately constant) when 
$H > \sqrt{V''(\phi)}$, and underdamped (and thus free to
roll) when $H < \sqrt{V''(\phi)}$.  The energy density is
$\rho_\phi = {1\over 2}\dot\phi^2 + V(\phi)$, and the
pressure is $p_\phi = {1\over 2}\dot\phi^2 - V(\phi)$,
implying an equation of state parameter
\begin{equation}
  w = {p\over \rho} = {{1\over 2}\dot\phi^2 - V(\phi) \over
  {1\over 2}\dot\phi^2 + V(\phi)}\ ,
\end{equation}
which will generally vary with time.  Thus, when the field
is slowly-varying and $\dot\phi^2 << V(\phi)$, we have
$w \sim -1$, and the scalar field potential acts like a
cosmological constant.

There are many reasons to consider dynamical dark energy
as an alternative to a cosmological constant.  First and foremost,
it is a logical possibility which might be correct, and can
be constrained by observation.  Secondly, it is consistent with
the hope that the ultimate vacuum energy might actually be zero, and
that we simply haven't relaxed all the way to the vacuum as yet.
But most interestingly, one might wonder whether replacing a constant
parameter $\Lambda$ with a dynamical field could allow us to
relieve some of the burden of fine-tuning that inevitably
accompanies the cosmological constant.  To date, investigations
have focused on scaling or tracker models of quintessence, in
which the scalar field energy density can parallel that of matter
or radiation, at least for part of its history 
\cite{Ferreira:1998,Copeland:1998,Zlatev:1999a,Liddle:1999,
Steinhardt:1999nw,Zlatev:1999b,Sahni:1999qe}.  
(Of course, we do not want the 
dark energy density to redshift away as rapidly as that in matter 
during the current epoch, or the universe would not be accelerating.)
Tracker models can be constructed in which the vacuum energy
density at late times is robust, in the sense that it does not
depend sensitively on the initial conditions for the field.
However, the ultimate value $\rho_{\rm vac}\sim (10^{-3} 
{\rm ~eV})^4$ still depends sensitively on the parameters in
the potential.  Indeed, it is hard to imagine how this could
help but be the case; unlike the case of the axion solution
to the strong-CP problem, we have no symmetry to appeal to that
would enforce a small vacuum energy, much less a particular
small nonzero number.

Quintessence models also introduce new naturalness problems
in addition to those of a cosmological constant.  These can be
traced to the fact that, in order for the field to be slowly-rolling
today, we require $\sqrt{V''(\phi_0)} \sim H_0$; but this 
expression is the effective mass of fluctuations in $\phi$, so
we have
\begin{equation}
  m_\phi \sim H_0 \sim 10^{-33} {\rm ~eV}\ .
\end{equation}
By particle-physics standards, this is an incredibly small
number; masses of scalar fields tend to be large in the
absence of a symmetry to protect them.  Scalars of such a low
mass give rise to long-range forces if they couple to ordinary
matter; since $\phi$ does couple to gravity, we expect at the
very least to have non-renormalizable interactions suppressed
by powers of the Planck scale.  Such interactions are potentially
observable, both via fifth-force experiments and searches for
time-dependence of the constants of nature, and current limits
imply that there must be suppression of the quintessence couplings
by several orders of magnitude over what would be expected
\cite{carroll,Chiba:1999xx,Horvat:1999}.  
The only known way to obtain such a suppression
is through the imposition of an approximate global symmetry
(which would also help explain the low mass of the field), of
the type characteristic of pseudo-Goldstone boson models of
quintessence, which have been actively explored
\cite{Frieman:1992,Frieman:1995,Kim:1999a,Choi:1999,Kim:1999b,
Nomura:1999}.  (Cosmological pseudo-Goldstone bosons are
potentially detectable through their tendency to rotate polarized
radiation from galaxies and the CMB \cite{carroll,Lue:1999}.)  
See \cite{Kolda:1998wq} for a discussion of further fine-tuning
problems in the context of supersymmetric models.

Nevertheless, these naturalness arguments are by no means
airtight, and it is worth considering specific particle-physics
models for the quintessence field.  In addition to the
pseudo-Goldstone boson models just mentioned, these include
models based on supersymmetric gauge theories
\cite{Bin:1999,Masiero:2000}, supergravity 
\cite{Brax:1999,Albrecht:1999}, small extra dimensions
\cite{Bento:1999,Barreiro:1999}, large extra dimensions
\cite{Benakli:1999,BDN:1999}, and non-minimal couplings to the 
curvature scalar \cite{Sahni:1998at,Uzan:1999,Amendola:1999,Parker:1999a,
Perrotta:1999,Perrotta:2000,deRitis:1999,Parker:1999b,Bertolami:1999}.
Finally, the possibility has been raised that the scalar field
responsible for driving inflation may also serve as quintessence
\cite{Frewin:1993,Peebles:1999,Peloso:1999,Giovannini:1999}, although
this proposal has been criticized for producing unwanted relics
and isocurvature fluctuations \cite{gkl}.

There are other models of dark energy besides those based
on nearly-massless scalar fields.  One scenario is ``solid''
dark matter, typically based on networks of tangled cosmic
strings or domain walls 
\cite{Vilenkin:1984,Spergel:1997,Bucher:1999,Battye:1999}.
Strings give an effective equation-of-state parameter
$w_{\rm string}=-1/3$, and walls have $w_{\rm wall}=-2/3$,
so walls are a better fit to the data at present.  There is
also the idea of dark matter particles whose masses increase
as the universe expands, their energy thus redshifting away more 
slowly than that of ordinary matter \cite{gb,ac} (see also
\cite{Hu:1998}).  The cosmological consequences of this kind of 
scenario turn out to be difficult to analyze analytically, and
work is still ongoing.

\section{Conclusions: the preposterous universe}

Observational evidence from a variety of sources currently
points to a universe which is (at least approximately) spatially
flat, with $(\Omega_{\rm M}, \Omega_\Lambda) \approx (0.3, 0.7)$.
The nucleosynthesis constraint implies that $\Omega_{\rm B}\sim
0.04$, so the majority of the matter content must be in an
unknown non-baryonic form.

\begin{figure}[t]
  \epsfxsize = 8cm
  \centerline{\epsfbox{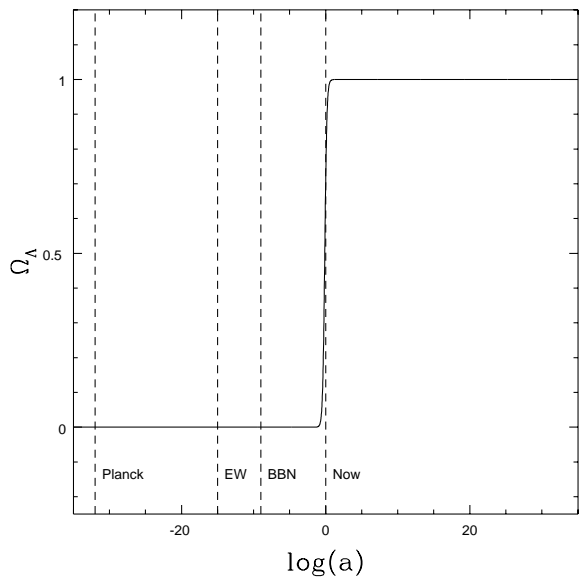}}
  \caption{$\Omega_\Lambda$ as a function of the scale factor $a$,
  for a universe in which $\Omega_{{\rm M}0}=0.3$, 
  $\Omega_{\Lambda 0}=0.7$.  Indicated are the scale factors 
  corresponding to the Planck era, the electroweak phase transition,
  and Big Bang Nucleosynthesis.}
  \label{omegalvsa}
\end{figure}

Nobody would have guessed that we live in such a universe.
Figure (\ref{omegalvsa}) is a plot of $\Omega_\Lambda$ as a 
function of the scale factor $a$ for this cosmology.  At early
times, the cosmological constant would have been negligible,
while at later times the density of matter will be 
essentially zero and the universe will be empty.  We happen
to live in that brief era, cosmologically speaking, when both
matter and vacuum are of comparable magnitude.  Within the
matter component, there are apparently contributions from
baryons and from a non-baryonic source, both of which are also
comparable (although at least their ratio is independent of
time).  This scenario staggers under the burden of its 
unnaturalness, but nevertheless crosses the finish line
well ahead of any competitors by agreeing so well with the
data.

Apart from confirming (or disproving) this picture, a major
challenge to cosmologists and physicists in the years to come
will be to understand whether these apparently distasteful
aspects of our universe are simply surprising coincidences,
or actually reflect a beautiful underlying structure we do
not as yet comprehend.  If we are fortunate, what appears 
unnatural at present will serve as a clue to a deeper
understanding of fundamental physics.

\section*{Acknowledgments}

I wish to thank Greg Anderson, Tom Banks, Robert Caldwell,
Gordon Chalmers, Michael Dine, George Field, Peter Garnavich, 
Jeff Harvey, Gordy Kane, Manoj Kaplinghat,
Bob Kirshner, Lloyd Knox, Finn Larsen, Laura Mersini, Ue-Li Pen,
Saul Perlmutter, Joe Polchinski, Ted Pyne, Brian Schmidt,
and Michael Turner for numerous useful conversations, 
Patrick Brady, Deryn Fogg and Clifford Johnson for rhetorical
encouragement, and Bill Press and Ed Turner for insinuating
me into this formerly-disreputable subject.

\end{document}